\documentstyle{article}

\newtheorem{Fact}{Fact}
\newtheorem{Proposition}{Proposition}

\newcommand{\subsupsize}[1]{\mbox{\scriptsize$#1$}}
\newcommand{\scriptbf}[1]{\mbox{\scriptsize${\bf #1}$}}

\newcommand{\scriptcal}[1]{\mbox{\scriptsize${\cal #1}$}}
\newcommand{\footnotecal}[1]{\mbox{\footnotesize${\cal #1}$}}

\newcommand{\smallsf}[1]{\mbox{\small${\sf #1}$}}

\newcommand{\minus}{\: \dot{-} \:}

\newcommand{\morph}{\! : \!}

\newcommand{\memberof}{\! \in \!}

\newcommand{\product}[2]{\mbox{$ #1 \! \times \! #2 $}}

\newcommand{\pair}[2]{\mbox{$ \langle #1,#2 \rangle $}}
\newcommand{\triple}[3]{\mbox{$ \langle #1,#2,#3 \rangle $}}
\newcommand{\quadruple}[4]{\mbox{$ \langle #1,#2,#3,#4 \rangle $}}
\newcommand{\quintuple}[5]{\mbox{$ \langle #1,#2,#3,#4,#5 \rangle $}}

\newcommand{\parpair}[4]
   {\; \left.
    \begin{picture}(20,12)(-10,-5)
       \put(-10,0){\makebox(0,0){$#1$}}
       \put(10,0){\makebox(0,0){$#4$}}
       \put(0,1.5){\makebox(0,0){$\rightharpoondown$}}
       \put(0,-3){\makebox(0,0){$\rightharpoondown$}}
       \put(0,7){\makebox(0,0){$\mbox{\small$#2$}$}}
       \put(0,-7){\makebox(0,0){$\mbox{\small$#3$}$}}
    \end{picture}
    \right. \;}

\newcommand{\bigparpair}[4]
   {\; \left.
    \begin{picture}(35,12)(-25,-5)
       \put(-20,0){\makebox(0,0){$#1$}}
       \put(10,0){\makebox(0,0){$#4$}}
       \put(0,1.5){\makebox(0,0){$\rightharpoondown$}}
       \put(0,-3){\makebox(0,0){$\rightharpoondown$}}
       \put(0,7){\makebox(0,0){$\mbox{\small$#2$}$}}
       \put(0,-7){\makebox(0,0){$\mbox{\small$#3$}$}}
    \end{picture}
    \right. \;}

\newcommand{\directflow}
   {\begin{picture}(8,8)(-4,-4)
       \put(0,0){\oval(8,8)[t]}
       \put(2,0){\oval(4,4)[b]}
       \put(-2,0){\oval(4,4)[t]}
    \end{picture}}

\newcommand{\inverseflow}
   {\begin{picture}(8,8)(-4,-4)
       \put(0,0){\oval(8,8)[b]}
       \put(2,0){\oval(4,4)[b]}
       \put(-2,0){\oval(4,4)[t]}
    \end{picture}}

\newcommand{\yiya}[1]
   {\setlength{\unitlength}{#1pt}
    \begin{picture}(1,1)(-.5,-.5)
       \put(0,0){\circle{1}}
       \put(.25,0){\oval(.5,.5)[b]}
       \put(-.25,0){\oval(.5,.5)[t]}
    \end{picture}
    \setlength{\unitlength}{1pt}}

\newcommand{\yinyang}{\yiya{8}}

\newcommand{\yayi}[1]
   {\setlength{\unitlength}{#1pt}
    \begin{picture}(1,1)(-.5,-.5)
       \put(0,0){\circle{1}}
       \put(.25,0){\oval(.5,.5)[t]}
       \put(-.25,0){\oval(.5,.5)[b]}
    \end{picture}
    \setlength{\unitlength}{1pt}}

\newcommand{\yangyin}{\yayi{8}}

\newcommand{\doublespace}{\vspace*{\baselineskip} \\}
\title{Introduction to Dialectical Nets \\
        $\pi \alpha \lambda \iota \nu \tau o \nu o \zeta$      $\alpha \rho \mu o \nu \iota \eta$
      - $\pi \alpha \lambda \iota \nu \tau \rho o \pi o \zeta$ $\alpha \rho \mu o \nu \iota \eta$}
\author{Robert E. Kent \\
        Department of Electrical Engineering and Computer Science \\
        University of Illinois at Chicago}

\begin{document}
   \begin{titlepage}

\noindent
{\bf INTRODUCTION TO DIALECTICAL NETS} \doublespace
\indent
ROBERT E. KENT \\
\indent
Department of Electrical Engineering and Computer Science \\
\indent
University of Illinois at Chicago, Chicago, Illinois \doublespace

\hspace*{\fill} ABSTRACT \hspace*{\fill} \doublespace
This paper initiates the dialectical approach to net theory.
This approach views nets as special,
but very important and natural,
dialectical systems.
By following this approach,
a suitably generalized version of nets,
called {\em dialectical nets},
can be defined in terms of the ``fundamental contradiction'' inherent in the structure of {\em closed preorders}.
Dialectical nets are the least conceptual upper bound
subsuming the notions of Petri nets, Kan quantification and transition systems.
The nature of dialectical nets is that of logical dynamics,
and is succinctly defined and summarized in the statement that
``dialectical nets are transition systems relativized to closed preorders,
and hence are general predicate transformers''. \doublespace


\hspace*{\fill} INTRODUCTION \hspace*{\fill} \doublespace
Nets are extensively used to model system phenomena such as concurrency, conflict, synchronization, information flow etc.
In order to model a variety of systems,
nets come in a variety of forms including:
condition/event nets using Boolean values,
consumption/production nets using natural numbers,
predicate/transition nets using colored tokens and formal polynomials,
etc.
In order to make mathematical sense out of this multiplicity of net models,
and in order to be able to extend the net concept to first order and higher order logic,
we stress the need for a proper mathematical base.
The first step in the development of this base
is recognition of the fact that values in nets
(be they booleans, numbers, colored tokens, subsets, etc.)
form a well-known mathematical structure called a {\em closed preorder}.
This is the established and accepted structure for predicates in first order and higher order logic
and should also be used in net theory.
The second step in the development of this base
is recognition of the fact that transitions in nets
(be they precondition/postcondition, consumption/production, conjunction/implication, existential-quantification/substitution, etc.)
are generalized inverse or dialectical activities
and form a well-known mathematical structure called an {\em adjunction}.
After development of a proper mathematical base for nets
we introduce a generalized model for such system phenomena called {\em dialectical nets}.
Dialectical nets are special, but very important, forms of dialectical systems based upon the internal contradiction in closed preorders.
The theory of dialectical (motion in) systems has already been applied to four important areas of computer science:
\begin{itemize}
\item concurrent systems \cite{Kent87a},
where it distinguishes the notions of observational equivalence and dialectical motion of transition systems;
\item OBJ-like functional programming,
where it generalizes the notion of {\em institutions},
and is based upon the doctrinal diagram associated with algebraic theories;
\item generalized Petri net theory [this paper],
where it unites notions of {\em nets} with the notion of {\em predicate transformers},
and is based upon the notions of bimodules, Kan quantification and normed categories; and
\item first order logic \cite{Kent87b},
where it unifies the semantics of Horn clause logic with that of relational databases,
and is based upon the notion of model doctrines.
\end{itemize}
The theory of dialectical systems was originally developed out of a desire to understand mathematically the obvious structural similarities between the "parallel composition" of concurrent systems and the "natural join" of database relations.
The dialectical view of nature \cite{Bernow&Raskin} is ancient:
it was discussed in the earliest history of ideas by various Presocratic Greek philosophers,
most notably Heraclitus.
Dialectical systems contain the following essential aspects:
1. based upon contradictions or "opposing tendencies";
2. interacting objects or entities;
3. movement, motion or development; and
4. reproduction or renewal of entities.
All of these aspects are present in the parallel composition of concurrent interacting systems and the natural join of database relations.
They are also present as basic concepts in the theory of nets.
F.W. Lawvere gave the theory of dialectical systems its most succinct expression:
{\sc Category Theory $\equiv$ Objective Dialectics}.
Indeed dialectics invests the dynamical view of systems theory with the fundamental ideas of category theory, such as adjunctions, limits, tensors and Kan extensions;
but in turn, it gives these categorical notions that dynamical view.
In short,
the theory of dialectics studies
both the ``motion (development, or growth) of structure''
and the ``structure of motion''. \doublespace

  \end{titlepage}
   \newpage

\hspace*{\fill} ENRICHED NETS \hspace*{\fill} \doublespace
In the theory of dialectics inverse activities such as consumption and
production are fundamental structural units called opposing tendencies, or contradictions.
A mathematical formulation of dialectics exists,
and is called category theory:
{\sc category theory $\equiv$ objective dialectics}.
In category theory dialectical contradictions are represented by
adjunctions.
Given two monotonic functions
${\cal B} \stackrel{f}{\longrightarrow} {\cal A}$ and
${\cal B} \stackrel{u}{\longleftarrow} {\cal A}$
flowing in opposite directions between two preorders
${\cal A} = \pair{A}{\preceq_{A}}$ and ${\cal B} = \pair{B}{\preceq_{B}}$,
the pair $\pair{f}{u}$ is called an
{\em adjunction} (or an adjoint pair, or a Galois connection, or generalized inverses, or opposing tendencies, or a dialectical contradiction),
when they satisfy
the equivalence axiom:
\( f(b) \preceq_{A} a \mbox{ iff } b \preceq_{B} u(a) \).
In this case,
$f$ is called the left adjoint (or left aspect) and $u$ is called the right adjoint (or right aspect).
The equivalence axiom can be interpreted as the ``dialectical tension''
which exists between the left and right aspects within the complementary pair.
We symbolize this adjunction by the functional notation
$(f \dashv u) \morph {\cal B} \longrightarrow {\cal A}$
with ${\cal B}$ (arbitrarily) the source of the adjunction and ${\cal A}$ the target.
The following two conditions are equivalent:
(1) $(f \dashv u) \morph {\cal B} \longrightarrow {\cal A}$;
(2) unit axiom $\mbox{Id}_{B} \preceq f \cdot u$;
that is, $b \preceq_{B} u(f(b))$ for all $b \in B$; and
counit axiom $u \cdot f \preceq \mbox{Id}_{A}$;
that is, $f(u(a)) \preceq_{A} a$ for all $a \in A$.
Either of these equivalent conditions implies the condition:
(3) $u \cdot f \cdot u \equiv u$; and
$f \cdot u \cdot f \equiv f$.
Ordinary inverses are generalized inverses (contradictions): 
two monotonic functions
${\cal B} \stackrel{f}{\longrightarrow} {\cal A}$ and ${\cal B} \stackrel{f^{-1}}{\longleftarrow} {\cal A}$
which are inverse to each other,
$f \cdot f^{-1} = {\rm Id}_B$ and $f^{-1} \cdot f = {\rm Id}_A$,
form an adjunction
$(f \dashv f^{-1}) \morph {\cal B} \longrightarrow {\cal A}$.
The notion of adjoint pairs can be generalized from the realm of preorders and monotonic functions to the realm of categories and functors.

The fundamental algebraic structure used to define the dynamics of consumption/production Petri nets
is that of the natural numbers ${\bf N}$.
Natural numbers represent quantities of various resources in systems,
which are distributed over, and indexed by, places.
Certain properties of natural numbers are essential in the definition of the structure and behavior of nets.
These properties form a coherent and very important mathematical structure called a closed preorder.
A {\em closed preorder}
\cite{Lawvere73}
${\bf V} = \quintuple{V}{\preceq}{\oplus}{\Rightarrow}{e}$
consist of the following data and axioms:
(1) $\quadruple{V}{\preceq}{\oplus}{e}$ is a monoidal preorder,
or ordered monoid,
with $\pair{V}{\preceq}$ a preorder and $\triple{V}{\oplus}{e}$ a monoid,
where the binary operation $\oplus \morph \product{V}{V} \longrightarrow V$, called {\bf V}-composition, is monotonic:
if both $u \preceq u'$ and $v \preceq v'$ then $(u \oplus v) \preceq (u' \oplus v')$;
(2) $\oplus$ is symmetric, or commutative; that is,
$a \oplus b = b \oplus a$ for all elements $a,b \in V$; and
(3) {\bf V} satisfies the closure axiom:
the monotonic {\bf V}-composition function
$(\:) \oplus b \morph V \longrightarrow V$
has a specified right adjoint
$b \!\Rightarrow\! (\:) \morph V \longrightarrow V$
for each element $b \in B$,
called {\bf V}-implication,
or symbolically
$\left( (\:) \oplus b \right) \dashv \left( b \!\Rightarrow\! (\:) \right) \morph V \longrightarrow V$;
that is,
$a \oplus b \preceq c$ iff $a \preceq b \!\Rightarrow\! c$
for any triple of elements $a,b,c \in V$.
The adjunction in the closure axiom is what we referred to as the fundamental contradiction inherent in the mathematical structure of the closed preorder {\bf V}.
The counit axiom for the closure adjunction is generalized {\em modus ponens}:
$((b \!\Rightarrow\! a) \oplus b) \preceq a$ for all elements $a,b \memberof V$.
When the unit axiom of the closure adjunction is equivalence,
$((b \!\Rightarrow\! (a \oplus b)) \equiv a$ for all elements $a,b \memberof V$,
the closed preorder {\bf V} is said to be {\em coreflective}.
The commutative, associative and unital binary operation $\oplus$ is sometimes called a tensor product.
We usually also assume that our closed preorders are bicomplete;
that is,
the supremum $\bigvee B$ and the infimum $\bigwedge B$ exist
(and are unique up to equivalence $\equiv$)
for all subsets $B \subseteq V$.
When the tensor product $\oplus$ is the binary infimum or meet $\wedge$
and the unit $e$ is the top element $\top_V$,
the closed preorder
${\bf V} = \quintuple{V}{\preceq}{\wedge}{\Rightarrow}{\top_V}$
is called a {\em cartesian closed preorder} (or a bicomplete Heyting prealgebra, or a locale).
The context of cartesian closed preorders is the context of traditional logic.
A characteristic property of cartesian closed preorders is idempotency:
$v \oplus v = v \wedge v = v$ for all elements $v \memberof V$.
In a cartesian closed preorder,
and even in an arbitrary closed preorder,
we regard $V$ as being a set of generalized truth values.
A closed preorder is normal when the unit is the top element $e = \top_V$
and {\bf V}-implication is directed-continuous:
$b \!\Rightarrow\!(\bigvee_{d \in D}d) \equiv \bigvee_{d \in D}(b \!\Rightarrow\! d)$ for all directed subsets $D \subseteq V$.
For normal closed preorders $a \oplus b \preceq a \wedge b$ for all elements $a,b \memberof V$.
Cartesian closed preorders are normal.

A pair ${\cal X} = \pair{X}{d_X}$ consisting of a set $X$ and a function
$d_X \morph \product{X}{X} \longrightarrow V$ is called a {\em quasi} {\bf V}-{\em space}
when it satisfies the triangle (or transitivity) axiom
$d_X(x_{1},x_{2}) \oplus d_X(x_{2},x_{3}) \preceq d_X(x_{1},x_{3})$
for all triples of elements $x_{1},x_{2},x_{3} \in X$;
and the zero (or reflexivity) axiom
$e \preceq d_X(x,x)$
for all elements $x \in X$.
The quasi {\bf V}-space ${\cal X} = \pair{X}{d_X}$ is a {\bf V}-space when it satisfies the additional condition:
if $e \preceq d_X(x_{1},x_{2})$ and $e \preceq d_X(x_{2},x_{1})$ then $x_{1} = x_{2}$.
The function $d_X$ is called a metric.
We interpret $d_X$ to be either a generalized distance function or a fuzzy preorder.
In general our metrics are asymmetrical:
$d_X(x_{1},x_{2}) \neq d_X(x_{2},x_{1})$.
Any quasi {\bf V}-space ${\cal X} = \pair{X}{d_X}$ can be symmetrized by defining
$d_X^{\rm sym}(x_{1},x_{2}) = d_X(x_{1},x_{2}) \oplus d_X^{\rm op}(x_{1},x_{2})$
where $d_X^{\rm op}(x_{1},x_{2}) = d_X(x_{2},x_{1})$ is the dual or opposite metric.
The set of truth values
${\cal V} = \pair{V}{d_V}$,
where $d_V(v_1,v_2) = v_1 \!\Rightarrow\! v_2$,
is a quasi {\bf V}-space.
Any set $X$ can be viewed as a discrete {\bf V}-space $X = \pair{X}{d_X} = X^{\rm op}$,
where $d_X(x,x') = e \mbox{ if } x \!=\! x', = \bot_V \mbox{ if } x \!\not=\! x'$.
Associated with every quasi {\bf V}-space
${\cal X} = \pair{X}{d_X}$
is an underlying preorder $\Box{\cal X} = \pair{X}{\preceq_X}$
where $x \preceq_X x'$ when $e \preceq d_X(x,x')$,
and $x$ and $x'$ are unrelated when $e \not\preceq d_X(x,x')$.
So the characteristic monotonic function for the order $\preceq_X$ is
$\kappa_{\preceq_X} = d_X \cdot \Box_V \morph \product{\pair{X}{\preceq_X}^{\rm op}}{\pair{X}{\preceq_X}} \rightarrow \pair{V}{\preceq} \rightarrow \pair{2}{\leq} = \{0\leq1\}$,
where $\Box_V = e \!\preceq\!(\:)$ is the usual characteristic function for the principal filter $\uparrow_V\!(e) \subseteq V$;
that is, $\Box_V(v) = 1$ if $e \preceq v$, and $\Box_V(v) = 0$ otherwise.
Note that $\Box({\cal X}^{\rm sym}) = (\Box{\cal X})^{\rm sym} = \pair{X}{\equiv_X}$
and that $\Box({\cal X}^{\rm op}) = (\Box{\cal X})^{\rm op} = \pair{X}{\succeq_X}$.
For a {\bf V}-space the underlying preorder is a partial order.
For a symmetric quasimetric {\bf V}-space the underlying preorder is an equivalence relation.
For the space of generalized truth values ${\cal V} = \pair{V}{d_V}$,
since $e \preceq d_V(v_{1},v_{2})$ iff $v_{1} \preceq v_{2}$,
the underlying preorder is the given order on ${\bf V}$.

A {\bf V}-{\em morphism} $f \morph {\cal X} \longrightarrow {\cal Y}$
between two quasi {\bf V}-spaces ${\cal X} = \pair{X}{d_X}$ and ${\cal Y} = \pair{Y}{d_Y}$
is a function $f \morph X \longrightarrow Y$ which satisfies the condition
$d_X(x,x') \preceq d_Y(f(x),f(x'))$ for all $x,x' \in X$.
{\bf V}-spaces and {\bf V}-morphisms form the category ${\bf Space}_V$.
By modus ponens,
$(\:) \oplus v \morph {\cal V} \longrightarrow {\cal V}$
is a {\bf V}-morphism for all elements $v \memberof V$.
By transitivity of $d_V$,
$v \!\Rightarrow\! (\:) \morph {\cal V} \longrightarrow {\cal V}$
is a {\bf V}-morphism for all elements $v \memberof V$.
Given any two quasi {\bf V}-spaces ${\cal X}$ and ${\cal Y}$ the set
of all {\bf V}-morphisms from ${\cal X}$ to ${\cal Y}$ is a quasi {\bf V}-space
${\cal Y}^{\cal X}$,
called the exponential quasi {\bf V}-space of $\cal X$ and ${\cal Y}$,
whose metric $d$,
called the pointwise inf metric,
is defined by
$d(f,g) = \bigwedge_{x \in X} d_Y(f(x),g(x))$.
Notice that the metric $d_X$ is not used to define $d$.
The metric $d_X$ is only used to restrict admission to the underlying set of ${\cal Y}^{\cal X}$.
In particular,
the exponential space ${\cal V}^{{\cal X}}$
of all {\bf V}-valued {\bf V}-morphisms on ${\cal X}$
is an quasi {\bf V}-space
with the inf metric
$d(\phi,\psi)
= \bigwedge_{x \in X} d_V(\phi(x),\psi(x))
= \bigwedge_{x \in X} [\phi(x) \!\Rightarrow\! \psi(x)]$.
We interpret an element of ${\cal V}^{{\cal X}}$,
a {\bf V}-morphism $\mu \morph {\cal X} \longrightarrow {\cal V}$,
to be an $X$-indexed marking $\mu \morph X \longrightarrow {\cal V}$
which satisfies the internal pointwise metric constraint $d_X$:
$d_X(x,x') \preceq d_V(\mu(x),\mu(x')) = \mu(x) \!\Rightarrow\! \mu(x')$ for all $x,x' \in X$;
or equivalently,
by the $\oplus$-$\Rightarrow$ adjunction,
$\mu(x) \oplus d_X(x,x') \preceq \mu(x')$ for all $x,x' \memberof X$.
Such a marking
$\mu \morph {\cal X} \rightarrow {\bf V}$
which is constrained by the metric $d_X$
is called a {\bf V}-({\em valued}) {\em predicate} over ${\cal X}$:
``predicates $\equiv$ metric-constrained markings''.
The specification
$\bot_V = d_X(x,x')$
is no constraint at all,
and that the specification
$e \preceq d_X(x,x')$
is precisely the order-theoretic constraint
$x \preceq x'$ requiring that $\mu$ satisfy
$\mu(x) \preceq \mu(x')$.
For the exponential space ${\cal V}^{{\cal X}}$
of {\bf V}-predicates over ${\cal X}$,
since
$e \preceq d(\phi,\psi)$
iff $e \preceq \bigwedge_{x \in X} d_V(\phi(x),\psi(x))$
iff $e \preceq d_V(\phi(x),\psi(x))$ for all $x \memberof X$
iff $\phi(x) \preceq \psi(x)$ for all $x \memberof X$,
the underlying preorder is the usual {\em entailment order} on {\bf V}-predicates over ${\cal X}$.

We relativize the notion of a consumption/production net
by using as our fundamental domain of values an arbitrary closed preorder {\bf V}
in place of the natural numbers {\bf N}.
A {\bf V}-{\em net} {\sf N} is a quadruple ${\sf N} = \quadruple{T}{{\cal P}}{\iota}{o}$ consisting of:
a set (of transition symbols) $T$,
a quasi {\bf V}-space (of places) ${\cal P}$,
an input weighting function $\iota \morph T \longrightarrow {\bf V}^{\cal P}$, and
an output weighting function $o \morph T \longrightarrow {\bf V}^{\cal P}$.
Markings are given two interpretations:
(1) a place $p$ is a site associated with a value (of a resource) $\mu(p)$
    and a marking $\mu$ is a distribution of places (hence resources); or
(2) a marking $\mu$ is a fuzzy $P$-subset with $\mu(p)$ indicating the degree-of-membership of ``$p \memberof \mu$''.
A transition $t \memberof T$ in a {\bf V}-net $\sf N$ is enabled by marking $\mu$ when $\mu \preceq \iota(t)$;
that is, when $\mu(p) \preceq \iota(t,p)$ for all places $p \memberof P$.
A transition $t \memberof T$ fires by $\Rightarrow$-ing, or consuming, $\iota(t,p)$ tokens from place $p \memberof P$,
and then $\oplus$-ing, or producing, $o(t,p)$ tokens to place $p \memberof P$.
The result of the firing of a transition $t \memberof T$ is expressed by the equation
${\sf N}_t(\mu)(p) =  [ \iota(t,p) \!\Rightarrow\! \mu(p) ] \oplus o(t,p)$
for all places $p \memberof P$.
We regard a {\bf V}-net to be a transformer of constrained {\bf V}-markings;
that is,
a {\bf V}-predicate transformer.
The semantics of a {\bf V}-net {\sf N} can be defined as either external or internal behaviors.
External behaviors include: (1) unfoldment-tree, and (2) regular-set behavior.
Internal behaviors include: (1) reachable predicates (markings), and (2) cumulative fixpoint behavior.

We list some important closed preorders on which nets and transition systems can be based: \\
{\bf booleans} [cartesian closed] \\
{\bf 2} = $\quintuple{2=\{0,1\}}{\leq}{\wedge}{\rightarrow}{1}$,
               where 0 is {\bf false}, 1 is {\bf true}, $\leq$ is the usual order on truth-values,
                     $\wedge$ is the truth-table for {\bf and}, and $\rightarrow$ is the truth-table for {\bf implies}.
               Here quasi {\bf 2}-spaces ${\cal X} = \pair{X}{d}$ are preorders ${\cal X} = \pair{X}{\preceq}$
                       where $x_1 \preceq x_2$ when $d(x_1,x_2) = 1$,
                    {\bf 2}-spaces are posets, and
                    {\bf 2}-morphisms are monotonic functions. \\
${\bf 2'}$ = $\quintuple{2=\{1,0\}}{\geq}{\vee}{\setminus}{0}$,
               where 1 is {\bf true}, 0 is {\bf false}, $\geq$ is the usual downward order on truth-values,
                     $\vee$ is the truth-table for {\bf or}, and $\setminus$ is the truth-table for {\bf difference}:
                        $b_1 \setminus b_2$ is true iff $b_1$ is true and $b_2$ is false.
               Here, quasi ${\bf 2'}$-spaces are preorders
                       where $x_1 \preceq x_2$ when $d(x_1,x_2) = 0$,
                     ${\bf 2'}$-spaces are posets, and
                     ${\bf 2'}$-morphisms are monotonic functions.
                     ${\bf 2'}$ defines the correct context for condition/event nets. \\
{\bf natural numbers} \\
      {\bf N} = $\quintuple{N}{\geq}{+}{\minus}{0}$,
where $N$ is the set of natural numbers $N = \{0,1,\ldots,n,\ldots,\infty\}$ with infinity,
$\geq$ is the usual downward ordering on natural numbers $N$,
$+$ is sum, and
$\minus$ is difference defined by
$m \minus n = m - n \mbox{ if } m \geq n, = 0 \mbox{ if } m < n$. \\
{\bf reals} \\
      {\bf R} = $\quintuple{R=[\infty,0]}{\geq}{+}{\minus}{0}$. \\
The quantitative closed preorders of reals {\bf R} and natural numbers {\bf N} are coreflective and normal.
They define the correct context for consumption/production nets. \\
{\bf markings} \\
      If {\bf V} is a closed preorder and $I$ is any indexing set,
      then the marking space ${\cal V}^I$ is a closed preorder
      ${\bf V}^I = \quintuple{V^I}{\preceq}{\oplus}{\Rightarrow}{e}$
      where $\preceq,\oplus,\Rightarrow \mbox{ and } e$ have obvious pointwise definitions.
      $I$ might denote places, colors, some combination of these, etc., in nets.
      If ${\bf V} = \quintuple{V}{\preceq}{\wedge}{\Rightarrow}{\top_V}$ is a cartesian closed preorder and ${\cal X}$ is any {\bf V}-space,
      then the predicate space ${\bf V}^{{\cal X}}$,
      the restriction of ${\bf V}^X$ to {\bf V}-morphisms,
      is a [cartesian closed] subpreorder of ${\bf V}^X$. \\
{\bf subsets} [cartesian closed] \\
      Let $A$ be any set and let $P(A)$ be the set $P(A)=\{B \mid B \subseteq A\}$ of all subsets of $A$. \\
${\bf P}(A) = \quintuple{P(A)}{\subseteq}{\cap}{\rightarrow}{A}$,
                 where $\cap$ is {\bf set intersection}, and $\rightarrow$ is {\bf set implication}:
                        $B_1 \rightarrow B_2 = \{a \memberof A \mid a \memberof B_1 \mbox{ implies } a \memberof B_2\} = -B_1 \cup B_2$.
               ${\bf P}(A)$ is essentially the marking space closed preorder ${\bf P}(A) \cong {\bf 2}^A$
               defining the most basic markings-as-fuzzy-subsets interpretation for nets. \\
${\bf P}'(A) = \quintuple{P(A)}{\supseteq}{\cup}{\setminus}{\emptyset}$,
               where $\cup$ is {\bf set union}, and $\setminus$ is {\bf set difference}:
                        $B_1 \setminus B_2 = \{a \memberof A \mid a \memberof B_1 \mbox{ but not } a \memberof B_2\} = B_1 \cap -B_2$.
               ${\bf P}'(A)$ is essentially the marking space closed preorder ${\bf P}'(A) \cong {\bf 2'}^A$,
               the marking space for condition/event nets. \\
{\bf propositional logic} [cartesian closed] \\
      Let $A$ be any fixed denumerable set of propositional variables
      and let ${\bf \Phi}(A)$ be the recursively defined set of all sentences.
      ${\bf \Phi}(A) = \quintuple{{\bf \Phi}(A)}{\models}{\wedge}{\rightarrow}{\top}$,
         where $\models$ is semantically defined logical entailment, and
         $\wedge$ and $\rightarrow$ are the syntactic binary operations on ${\bf \Phi}(A)$
         defined by $\wedge(\alpha,\beta) = (\alpha \wedge \beta)$
                and $\rightarrow(\alpha,\beta) = (\alpha \rightarrow \beta)$.
      Here, quasi ${\bf \Phi}(A)$-spaces are sentence-valued sets.
      A ${\bf \Phi}(A)$-marking $\mu$ assigns a sentence $\mu(p)$ to each place $p \memberof P$,
      hence is a $P$-indexed collection of sentences.
      The metric $d_P$,
      in the quasi ${\bf \Phi}(A)$-space of places ${\cal P} = \pair{P}{d_P}$,
      specifies generalized laws of modus ponens,
      since if $d_P(p_1,p_2) = \alpha$ then $\alpha \models \mu(p_1) \rightarrow \mu(p_2)$;
      that is, $\alpha \wedge \mu(p_1) \models \mu(p_2)$.
      So for all ordered pairs $(p_1,p_2)$ of places
      $d_P$ specifies an assumption, or context, in which
      $\mu(p_1)$ the sentence indexed at place $p_1$
      is required to logically entail
      $\mu(p_2)$ the sentence indexed at place $p_2$.
      In particular, $d_P(p_1,p_2)$ = ``$\mu(p_1) \rightarrow \mu(p_2)$''
      specifies ordinary modus ponens,
      $(\mu(p_1) \rightarrow \mu(p_2)) \wedge \mu(p_1) \models \mu(p_2)$,
      the weakest assumption for logical entailment between $\mu(p_1)$ and $\mu(p_2)$.

There is a standard method for transforming between enriched contexts.
A closed monotonic function $h \morph {\bf V} \longrightarrow {\bf W}$
between two closed preorders ${\bf V}$ and ${\scriptbf{W}}$
is a monotonic function $h \morph {\cal V} \longrightarrow {\cal W}$
which preserves monoidal unit and composition:
$e_{\scriptbf{W}} \preceq_{\scriptbf{W}} h(e_{\scriptbf{V}})$ and
$h(v) \oplus_{\scriptbf{W}} h(v') \preceq_{\scriptbf{W}} h(v \oplus_{\scriptbf{V}} v')$
for all elements $v,v' \memberof V$.
Closed monotonic functions determine transformations of spaces, predicates, relations, dialectical moves, dialectical nets, etc.
For example,
any closed preorder {\bf V} has a canonical closed monotonic function
$\Box_V \morph {\bf V} \longrightarrow {\bf 2}$,
defined above,
which determines a functor
$\Box_V \morph {\bf Space}_V \longrightarrow {\bf Space}_2\!=\!{\bf PO}$,
where $\Box_V({\cal X}) = \pair{X}{\preceq_X}$ the underlying preorder of ${\cal X}$.
Some simple net oriented examples:
(1) the two monotonic functions,
inclusion ${\rm Inc} \morph {\bf N} \longrightarrow {\bf Z}$ and
saturation $[\:] \morph {\bf Z} \longrightarrow {\bf N}$,
between natural numbers {\bf N} and integers {\bf Z},
where $[n] = n \mbox{ if } n \geq 0, = 0 \mbox{ if } n < 0$,
are adjoint $({\rm Inc} \dashv [\:])$ and closed
(allowing the use of linear algebraic techniques in net theory);
(2) the two monotonic functions,
floor $\lfloor\:\rfloor \morph {\bf R} \longrightarrow {\bf N}$ and
inclusion ${\rm Inc} \morph {\bf N} \longrightarrow {\bf R}$,
between (nonnegative) reals {\bf R} and natural numbers {\bf N},
are adjoint $(\lfloor\:\rfloor \dashv {\rm Inc})$ and closed
(encouraging the use of analysis techniques in net theory).


If ${\bf V} = \quintuple{V}{\preceq}{\oplus}{\Rightarrow}{e}$ is not a cartesian closed preorder,
such as {\bf N},
then ${\bf V}^{{\cal X}}$ is not necessarily a closed preorder
since it may not be closed under the pointwise operations $\oplus$ and $\Rightarrow$.
For a counterexample,
let ${\cal X}$ be the symmetric two point {\bf N}-space
${\cal X} = \pair{X}{d}$
where $X=\{a,b\}$ and $d(a,b) = 1 = d(b,a)$,
and let $\phi \morph {\cal X} \longrightarrow {\cal N}$ be defined by $\phi(a)=1$ and $\phi(b)=2$.
If $\theta \morph {\cal X} \longrightarrow {\cal N}$ is defined by $\theta(a)=1$ and $\theta(b)=0$,
then $\phi \minus \theta$,
where $(\phi \minus \theta)(a)=0$ and $(\phi \minus \theta)(b)=2$,
is not an {\bf N}-morphism ${\cal X} \longrightarrow {\cal N}$
since $d(b,a) \not\geq (\phi \minus \theta)(b) \minus (\phi \minus \theta)(a)$.
In the other dialectical direction,
if $\theta \morph {\cal X} \longrightarrow {\cal N}$ is defined to be $\phi$ above,
then $\phi+\theta = \phi+\phi = 2\phi$,
where $(\phi+\theta)(a)=2\phi(a)=2$ and $(\phi+\theta)(b)=2\phi(b)=4$,
is not an {\bf N}-morphism ${\cal X} \longrightarrow {\cal N}$
since $d(b,a) \not\geq (\phi + \theta)(b) \!\Rightarrow\! (\phi + \theta)(a)$.
In one sense the problem with the pointwise and discrete operations of implication $\theta \!\Rightarrow\! (\:)$ and composition $(\:) \oplus \theta$ 
is that they are ``isolated'' notions without ``collective'' influence between points,
whereas the metric $d$ makes $X$ into a nondiscrete structure ${\cal X} = \pair{X}{d}$
with points which are not isolated from one another in the sense that they have collective constraints between themselves.
One general solution to this problem is the use of enriched relations to model the dialectical movement of consumption and production.
Relations allow for collective influence between points. \doublespace


\hspace*{\fill} DIALECTICAL NETS \hspace*{\fill} \doublespace
Each element $x \memberof X$ of a quasi {\bf V}-space ${\cal X} = \pair{X}{d}$
can be represented as the {\bf V}-predicate ${\rm y}(x) = d(x,-)$ over ${\cal X}$
where ${\rm y}(x)(x') = d(x,x')$ for each element $x' \memberof X$.
The map ${\rm y} \morph X \longrightarrow V^X$,
which is called the {\em Yoneda embedding},
is a {\bf V}-isometry ${\rm y}_{{\cal X}} \morph {\cal X}^{\rm op} \longrightarrow {\bf V}^{\cal X}$.
Composition on the right with the Yoneda embedding
allows us to consider the concept of a {\bf V}-morphism
${\cal Y}^{\rm op} \longrightarrow {\bf V}^{\cal X}$
to be a generalization of the concept of a {\bf V}-morphism
${\cal Y} \longrightarrow {\cal X}$. 
Such a generalized {\bf V}-morphism is equivalent to a {\bf V}-morphism
$\product{{\cal Y}^{\rm op}}{{\cal X}} \stackrel{\tau}{\longrightarrow} {\bf V}$
and may be regarded to be a {\bf V}-({\em valued}) {\em relation}
(also called a {\bf V}-{\em bimodule})
from {\cal Y} to {\cal X},
denoted by ${\cal Y} \stackrel{\tau}{\rightharpoondown} {\cal X}$,
with $\tau(y,x)$ a element of {\bf V} being interpreted as the
``truth-value of the $\tau$-relatedness of $y$ to $x$'' \cite{Lawvere73}.
We can regard a {\bf V}-relation to be a
$\product{|{\cal Y}|}{|{\cal X}|}$-matrix
whose $(y,x)$-th entry is $\tau(y,x)$.
As mentioned above
every {\bf V}-morphism ${\cal Y} \stackrel{f}{\longrightarrow} {\cal X}$
determines a {\bf V}-relation ${\cal Y} \stackrel{f^\triangleleft}{\rightharpoondown} {\cal X}$
defined by $f^\triangleleft = f^{\rm op} \cdot {\rm y}_{\cal X}$;
that is,
$f^\triangleleft(y,x) = d_X(f(y),x)$.
Dually every {\bf V}-morphism ${\cal Y} \stackrel{f}{\rightarrow} {\cal X}$ also determines a {\bf V}-relation ${\cal X} \stackrel{f_\triangleleft}{\rightharpoondown} {\cal Y}$
defined by $f_\triangleleft = {\rm y}_{\cal X} \cdot {\bf V}^f$;
that is,
$f_\triangleleft(x,y) = d_X(x,f(y))$.
A pair of {\bf V}-relations ${\cal Z} \stackrel{\sigma}{\rightharpoondown} {\cal Y}$ and ${\cal Y} \stackrel{\rho}{\rightharpoondown} {\cal X}$
can be composed,
yielding the {\bf V}-relation ${\cal Z} \stackrel{\sigma\circ\rho}{\rightharpoondown} {\cal X}$
defined to be the categorical {\em coend}
$\sigma\!\circ\!\rho(z,x)
= \int^{y \in {\cal Y}} [\sigma(z,y) \oplus \rho(y,x)]
= \bigvee_{y \in {\cal Y}} [\sigma(z,y) \oplus \rho(y,x)]$
where $\bigvee$ denotes supremum,
which is colimit,
in ${\bf V} = \pair{V}{\preceq}$.
Relational composition can be viewed as a matrix product.
One can verify that relational composition is associative
$(\tau\circ\sigma)\circ\rho = \tau\circ(\sigma\circ\rho)$,
and that metrics (as {\bf V}-relations) are identities
$d_Y \circ \tau = \tau = \tau \circ d_X$.
So {\bf V}-spaces and {\bf V}-relations form a category ${\bf rel}_{\scriptbf{V}}$.
One can also verify that
$ (g \cdot f)^\triangleleft = g^\triangleleft \circ f^\triangleleft $
for any two composable {\bf V}-morphisms ${\cal Z} \stackrel{g}{\rightarrow} {\cal Y} \stackrel{f}{\rightarrow} {\cal X}$,
and that $(\mbox{Id}_{\cal X})^\triangleleft = d_X$ the identity {\bf V}-relation at ${\cal X}$.
So the Yoneda embedding determines a functor
$(\:)^\triangleleft \morph {\bf Space}_{\scriptbf{V}} \longrightarrow {\bf rel}_{\scriptbf{V}}$
which makes concrete the concept generalization discussed at the beginning of this section.

There is a concept orthogonal to relations-as-morphisms.
Given any two {\bf V}-relations
${\cal Y}_1 \stackrel{\tau_1}{\rightharpoondown} {\cal X}_1$ and
${\cal Y}_2 \stackrel{\tau_2}{\rightharpoondown} {\cal X}_2$
a ({\em vertical}) {\em morphism} of {\bf V}-relations
$\pair{g}{f} \morph \triple{{\cal Y}_1}{\tau_1}{{\cal X}_1} \Rightarrow \triple{{\cal Y}_2}{\tau_2}{{\cal X}_2}$
consists of two {\bf V}-morphisms,
a source morphism $g \morph {\cal Y}_1 \longrightarrow {\cal Y}_2$ and
a target morphism $f \morph {\cal X}_1 \longrightarrow {\cal X}_2$,
which satisfy the inequality condition
$\tau_1 \preceq (\product{g^{\rm op}}{f}) \cdot \tau_2$
as {\bf V}-predicates over $\product{{\cal Y}_1^{\rm op}}{{\cal X}_1}$;
or more abstractly,
in terms of relational composition,
$g_\triangleleft \circ \tau_1 \preceq \tau_2 \circ f_\triangleleft$.
{\bf V}-relations and their vertical morphisms form a category ${\bf Rel}_{\scriptbf{V}}$.
This is actually the vertical category of a double category,
which we also denote by ${\bf Rel}_{\scriptbf{V}}$,
whose underlying horizontal category is ${\bf rel}_{\scriptbf{V}}$.
If we change the definition of a vertical morphism
$\pair{f}{g} \morph \triple{{\cal Y}_1}{\tau_1}{{\cal X}_1} \Rightarrow \triple{{\cal Y}_2}{\tau_2}{{\cal X}_2}$
to the inequality condition
$g^\triangleleft \circ \tau_2 \preceq \tau_1 \circ f^\triangleleft$,
we can define a dual (vertical) category,
which we denote by ${\bf Rel}^\bullet_{\scriptbf{V}}$.
For any category ${\bf C}$ the category of parallel pairs of {\bf C}-morphisms,
denoted by ${\bf C}^\times\!$,
has the same objects as ${\bf C}$, $|{\bf C}^\times\!| = |{\bf C}|$,
and has parallel pairs of ${\bf C}$-morphism as its morphisms, ${\bf C}^\times\![c,c'] = ({\bf C}[c,c'])^2$.
${\bf C}^\times\!$ is a kind of 2nd power (square) of ${\bf C}$.
In particular,
${\bf rel}_{\scriptbf{V}}^\times$ is the category of parallel relation pairs
$\tau = \parpair{{\cal Y}}{o}{\iota}{{\cal X}}$
with (horizontal) ${\bf rel}_{\scriptbf{V}}$-composition.
Then ${\bf rel}_{\scriptbf{V}}^\times$,
with the vertical morphisms
$\pair{g}{f} \morph \triple{{\cal Y}_1}{\tau_1}{{\cal X}_1} \Rightarrow \triple{{\cal Y}_2}{\tau_2}{{\cal X}_2}$ 
when
$\pair{g}{f} \morph \triple{{\cal Y}_1}{o_1}{{\cal X}_1} \Rightarrow \triple{{\cal Y}_2}{o_2}{{\cal X}_2}$ 
is a vertical morphism in ${\bf Rel}_{\scriptbf{V}}$
and
$\pair{f}{g} \morph \triple{{\cal Y}_1}{\iota_1}{{\cal X}_1} \Rightarrow \triple{{\cal Y}_2}{\iota_2}{{\cal X}_2}$
is a vertical morphism in ${\bf Rel}_{\scriptbf{V}}^\bullet$,
is a vertical category denoted ${\bf Rel}_{\scriptbf{V}}^\times$.

Any element $v \memberof V$ is a {\bf V}-relation
${\bf 1} \stackrel{v}{\rightharpoondown} {\bf 1}$.
In fact,
${\bf rel}_{\scriptbf{V}}[{\bf 1},{\bf 1}] = {\bf V}^{\product{{\bf 1}^{\rm op}}{{\bf 1}}} \cong {\bf V}$.
Any {\bf V}-relation ${\bf 1} \stackrel{\psi}{\rightharpoondown} {\cal Y}$,
a {\em generalized element} of ${\cal Y}$,
is a {\bf V}-morphism $\product{{\bf 1}^{\rm op}}{{\cal Y}} \stackrel{\psi}{\longrightarrow} {\bf V}$,
and hence is the same thing as a predicate over ${\cal Y}$:
``${\cal Y}$-predicates $\equiv$ generalized ${\cal Y}$-elements''.
In fact,
${\bf rel}_{\scriptbf{V}}[{\bf 1},{\cal Y}] = {\bf V}^{\product{{\bf 1}^{\rm op}}{{\cal Y}}} \cong {\bf V}^{{\cal Y}}$.
So any {\bf V}-relation ${\cal Y} \stackrel{\tau}{\rightharpoondown} {\cal X}$
defines by composition a {\bf V}-morphism
${\bf V}^{{\cal Y}} \! \stackrel{\directflow_\tau}{\longrightarrow} {\bf V}^{{\cal X}}$
called {\em direct flow} (or {\em yang}),
which maps {\bf V}-predicates over ${\cal Y}$ to {\bf V}-predicates over ${\cal X}$,
and is defined by
$\directflow_\tau\!(\psi) = \psi \circ \tau$.
The direct flow $\directflow_\tau$ corresponds to the direct image map $PY \stackrel{R_P}{\longrightarrow} PX$
of a ordinary binary relation $R \subseteq \product{Y}{X}$.
Explicitly,
$\directflow_\tau$ is defined to be the coend
\[ \directflow_\tau(\psi)(x)
= \int^{y \in {\cal Y}} [\psi(y) \oplus \tau(y,x)]
= \bigvee_{y \in {\cal Y}} [\psi(y) \oplus \tau(y,x)] \]
for any predicate $\psi \memberof {\bf V}^{{\cal Y}}$ over ${\cal Y}$ and any point $x \memberof {\cal X}$.
Note that
$\directflow_{\sigma \circ \rho} = \directflow_\sigma \cdot \directflow_\rho$
for any pair of composable {\bf V}-relations ${\cal Z} \stackrel{\sigma}{\rightharpoondown} {\cal Y}$ and ${\cal Y} \stackrel{\rho}{\rightharpoondown} {\cal X}$,
and that
$\tau_1 \preceq \tau_2$ implies $\directflow_{\tau_1} \preceq \directflow_{\tau_2}$ as {\bf V}-morphisms for all pairs
$\parpair{{\cal Y}}{\tau_1}{\tau_2}{{\cal X}}$.
For the special case ${\cal Y} = {\bf 1} = {\cal X}$,
when {\bf V}-elements are viewed as {\bf V}-relations
${\bf 1} \stackrel{v}{\rightharpoondown} {\bf 1}$,
direct flow is
$\directflow_v = (\:) \oplus v \morph {\bf V}\!=\!{\bf V}^{\bf 1} \longrightarrow {\bf V}^{\bf 1}\!=\!{\bf V}$
since
$\directflow_v(u) = \bigvee_{{\bf 1}} (u \oplus v) = u \oplus v$.
For any {\bf V}-morphism ${\cal Y} \stackrel{f}{\longrightarrow} {\cal X}$
with associated {\bf V}-relation ${\cal X} \stackrel{f_\triangleleft}{\rightharpoondown} {\cal Y}$,
note that
$\directflow_{f_\triangleleft}(\phi) = f \cdot \phi$
for any {\bf V}-predicate $\phi \memberof {\bf V}^{{\cal X}}$ over ${\cal X}$;
that is,
$\directflow_{f_\triangleleft}$ is just composition on the right $\directflow_{f_\triangleleft} = f \cdot (\:) = {\bf V}^f$.
On the other hand,
for any {\bf V}-morphism ${\cal Y} \stackrel{f}{\longrightarrow} {\cal X}$
with associated {\bf V}-relation ${\cal Y} \stackrel{f^\triangleleft}{\rightharpoondown} {\cal X}$,
the {\bf V}-morphism
$\exists_f = \directflow_{f^\triangleleft} \morph {\bf V}^{{\cal Y}} \! \longrightarrow {\bf V}^{{\cal X}}$
is called {\em existential Kan quantification} along $f$;
in detail,
$\exists_f(\psi)(x) = \bigvee_{y \in {\cal Y}} [\psi(y) \oplus d_X(f(y),x)]$
for any predicate $\psi \memberof {\bf V}^{{\cal Y}}$ over ${\cal Y}$ and any point $x \memberof {\cal X}$.
Applying the directflow operator to the inequality condition for vertical morphisms in ${\bf Rel}_{\scriptbf{V}}$ gives
${\bf V}^g \cdot \directflow_{\tau_1} \preceq \directflow_{\tau_2} \cdot {\bf V}^f$.

The {\bf V}-morphism $\directflow_\tau$ has as a right adjoint the {\bf V}-morphism
${\bf V}^{{\cal Y}} \! \stackrel{\inverseflow_\tau}{\longleftarrow} {\bf V}^{{\cal X}}$
called {\em inverse flow} (or {\em yin}),
which maps {\bf V}-predicates over ${\cal X}$ to {\bf V}-predicates over ${\cal Y}$,
and is defined to be the categorical {\em end}
\[ \inverseflow_\tau(\phi)(y)
= \int_{x \in {\cal X}} [\tau(y,x) \!\Rightarrow\! \phi(x)]
= \bigwedge_{x \in {\cal X}} [\tau(y,x) \!\Rightarrow\! \phi(x)]
= \bigwedge_{x \in {\cal X}} d_{\scriptbf{V}}(\tau(y,x),\phi(x)) \]
for any predicate $\phi \memberof {\bf V}^{{\cal X}}$ over ${\cal X}$ and any point $y \memberof {\cal Y}$,
where $\bigwedge$ denotes infimum,
which is limit,
in ${\bf V} = \pair{V}{\preceq}$.
Note that
$\inverseflow_{\sigma \circ \rho} = \inverseflow_\rho \cdot \inverseflow_\sigma$
for any pair of composable {\bf V}-relations ${\cal Z} \stackrel{\sigma}{\rightharpoondown} {\cal Y}$ and ${\cal Y} \stackrel{\rho}{\rightharpoondown} {\cal X}$,
and that
$\tau_1 \preceq \tau_2$ implies $\inverseflow_{\tau_2} \preceq \inverseflow_{\tau_1}$ as {\bf V}-morphisms for all pairs
$\parpair{{\cal Y}}{\tau_1}{\tau_2}{{\cal X}}$.
For the special case ${\cal Y} = {\bf 1} = {\cal X}$,
when {\bf V}-elements are viewed as {\bf V}-relations
${\bf 1} \stackrel{v}{\rightharpoondown} {\bf 1}$,
inverse flow is
$\inverseflow_v = v \!\Rightarrow\! (\:) \morph {\bf V}\!=\!{\bf V}^{\bf 1} \longrightarrow {\bf V}^{\bf 1}\!=\!{\bf V}$
since
$\inverseflow_v(u) = \bigwedge_{{\bf 1}} (v \!\Rightarrow\! u) = v \!\Rightarrow\! u$.
For any {\bf V}-morphism ${\cal Y} \stackrel{f}{\longrightarrow} {\cal X}$
with associated {\bf V}-relation ${\cal X} \stackrel{f^\triangleleft}{\rightharpoondown} {\cal Y}$,
note that
$\inverseflow_{f^\triangleleft}(\phi) = f \cdot \phi$
for any {\bf V}-predicate $\phi \memberof {\bf V}^{{\cal X}}$ over ${\cal X}$;
that is,
$\inverseflow_{f^\triangleleft}$ is just composition on the right
$\inverseflow_{f^\triangleleft} = f \cdot (\:) = {\bf V}^f = \directflow_{f_\triangleleft}$.
On the other hand for any {\bf V}-morphism ${\cal Y} \stackrel{f}{\longrightarrow} {\cal X}$
with associated {\bf V}-relation ${\cal X} \stackrel{f_\triangleleft}{\rightharpoondown} {\cal Y}$,
the {\bf V}-morphism
$\forall_f = \inverseflow_{f_\triangleleft} \morph {\bf V}^{{\cal Y}} \! \longrightarrow {\bf V}^{{\cal X}}$
is called {\em universal Kan quantification} along $f$;
in detail,
$\forall_f(\psi)(x) = \bigwedge_{y \in {\cal Y}} [d_X(x,f(y)) \!\Rightarrow\! \psi(y)]$
for any predicate $\psi \memberof {\bf V}^{{\cal Y}}$ over ${\cal Y}$ and any point $x \memberof {\cal X}$.
Ordinary quantification in predicate calculus is a very special case of Kan quantifiction.
Kan quantification is quantification relativized to an arbitrary closed preorder {\bf V}.
Applying the inverseflow operator to the inequality condition for vertical morphisms in ${\bf Rel}^\bullet_{\scriptbf{V}}$ gives
${\bf V}^f \cdot \inverseflow_{\tau_1} \preceq \inverseflow_{\tau_2} \cdot {\bf V}^g$.
The following fact appears in \cite{Lawvere73},
showing that (parallel pairs of) relations specify ``the dialectical flow of predicates''.
\begin{Fact}
Direct flow is left adjoint to inverse flow
$\left( \directflow_\tau \dashv \inverseflow_\tau \right)$
for any {\bf V}-relation ${\cal Y} \stackrel{\tau}{\rightharpoondown} {\cal X}$.
\end{Fact}

Dialectical flow is the alternation-composition of direct and inverse flow.
A parallel pair
$\tau = \parpair{{\cal Y}}{o}{\iota}{{\cal X}}$
of {\bf V}-relations is called a {\em dialectical flow specifier}.
Given a flow specifier $\tau$,
the {\em dialectical flow} (or {\em yinyang}) specified by $\tau$ is the composition
\[ \yinyang_\tau = {\bf V}^{\cal X} \stackrel{\inverseflow_\iota}{\longrightarrow} {\bf V}^{\cal Y} \stackrel{\directflow_o}{\longrightarrow} {\bf V}^{\cal X} .\]
The symbol $\yinyang$ denotes dialectical motion.
This is a stylized version of the {\em yin-yang} symbol of the ancient Chinese {\em naturalist} philosophers,
where the two parts {\em yin} and {\em yang} represent contradictory elements or opposing tendencies forming a complementary union out of which all things develop.
For us,
the complementary union is the flow adjunction
and the circular shape represents the cyclical or spiral shape of dialectical motion.
Dialectical flow is functorial:
if $\pair{g}{f} \morph \triple{{\cal Y}_1}{\tau_1}{{\cal X}_1} \Rightarrow \triple{{\cal Y}_2}{\tau_2}{{\cal X}_2}$ 
is a vertical morphism in ${\bf Rel}_{\scriptbf{V}}^\times$,
then ${\bf V}^f \cdot \yinyang_{\tau_1} \preceq \yinyang_{\tau_2} \cdot {\bf V}^f$.
There is a four-fold duality in the definition of dialectical flow.
The {\em dialectical opflow} (or {\em yangyin}) specified by $\tau$ is the composition
$\yangyin_\tau = \directflow_o \cdot \inverseflow_\iota$.
The \mbox{{\em dialectical}} {\em coflow} of $\tau$ is the dialectical flow of $\tau^{\rm op}$; namely,
$\yinyang_{\tau^{\rm op}} = \inverseflow_o \cdot \directflow_\iota$.
The {\em dialectical coopflow} of $\tau$ is the dialectical opflow of $\tau^{\rm op}$; namely,
$\yangyin_{\tau^{\rm op}} = \directflow_\iota \cdot \inverseflow_o$.

The modern theory of dialectics incorporates the notion of the {\em reproduction} of entities which are in dialectical motion.
Reproduction can be modelled mathematically as the recursive specification of entities with respect to dialectical motion.
Reproduction (renewal, recursion) is the internal semantics of dialectical motion.
Dialectical entities in the context of this paper are the enriched predicates (constrained markings).
For any flow specifier $\tau$, as above,
we say that $\tau$ \mbox{{\em reproduces}} the {\bf V}-predicate $\phi \memberof {\bf V}^{\cal X}$ when $\phi$ is a fixpoint solution of the recursive equation
\[ \chi \equiv \yinyang_\tau(\chi) ;\]
that is,
when $\phi \equiv \yinyang_\tau(\phi) = \directflow_o(\inverseflow_\iota(\phi))$.
A fixpoint solution $\phi$ is an internal behavior of the dialectical motion (specified by) $\tau$.
A measure, or index, of reproduction is the value
$|\phi|_\tau
= \yinyang_\tau^\triangleleft(\phi)
= d^{\rm sym}(\phi,\yinyang_\tau(\phi))$.
So the dialectical flow $\tau$ reproduces the entity ({\bf V}-predicate) $\phi$
iff $e \preceq |\phi|_\tau$.
The increasing sequence 
$\mbox{$\bot_{\cal X} \preceq \yinyang_\tau(\bot_{\cal X}) \preceq \cdots \preceq \yinyang_\tau^n(\bot_{\cal X}) \preceq \cdots$}$
of {\bf V}-predicates over quasi {\bf V}-space ${\cal X}$,
is called the least fixpoint approximation sequence of $\tau$.
The {\bf V}-predicate 
$\tau^\ast = \bigvee_{n \in \omega}\yinyang_\tau^n(\bot_{\cal X})$,
the {\em least fixpoint} solution of the recursive equation above,
exists for directed-continuous inverse flow (yin),
and in particular,
for normal {\bf V}.
The decreasing sequence 
$\mbox{$\cdots \preceq \yinyang_\tau^n(\top_{\cal X}) \preceq \cdots \preceq \yinyang_\tau(\top_{\cal X}) \preceq \top_{\cal X}$}$
of {\bf V}-predicates over quasi {\bf V}-space ${\cal X}$,
is called the greatest fixpoint approximation sequence of $\tau$.
The {\bf V}-predicate 
$\tau^\infty = \bigwedge_{n \in \omega}\yinyang_\tau^n(\top_{\cal X})$,
is the {\em greatest fixpoint} solution of the recursive equation above.
The least and greatest fixpoints are two canonical internal behaviors of the dialectical motion $\tau$.
[A philosophical note:
The notion of complementary union
(two working together in one)
is not that of ``synthesis''.
Neither of the opposites is ``transformed''.
Indeed,
with synthesis,
dialectical motion would cease! 
The notion of ``reproduction'' is one of equilibrium of motion,
not lack of motion.]


Just as in the ordinary context of sets a subset $A \subseteq X$ can be viewed as a binary relation $A \subseteq \product{X}{X}$ ($ yAx \mbox{ iff } x \memberof A$, or $R_A = \mbox{pr}_2 \cdot \kappa_A$),
so also in the context of {\bf V}-spaces a marking $\theta \morph X \longrightarrow V$ can be viewed as a {\bf V}-relation
${\cal X} \stackrel{\tau_\theta}{\rightharpoondown} \tilde{\cal X}$
for any two quasi {\bf V}-spaces ${\cal X} = \pair{X}{d}$ and $\tilde{\cal X} = \pair{X}{\tilde{d}}$.
In this spirit and following the places-as-sites interpretation of nets,
for any two quasi {\bf V}-spaces ${\cal X}$ and $\tilde{\cal X}$,
and any {\bf V}-relation ${\cal X} \stackrel{\tau}{\rightharpoondown} \tilde{\cal X}$,
we view inverse dialectical flow $\inverseflow_\tau$ as a generalization of the concept of consumption,
and direct dialectical flow $\directflow_\tau$ as a generalization of the concept of production.
\begin{Proposition}
Consumption is a special case of inverse dialectical flow.
In particular,
if the map
$\tau_{\theta} = (\mbox{pr}_2 \cdot \theta) \oplus d^{\rm sym} \oplus \tilde{d}$
associated with any marking $\theta$ over $X$ is a {\bf V}-relation,
then ordinary consumption by $\theta$ is inverse dialectical flow along $\tau_{\theta}$;
that is,
$\theta \!\Rightarrow\! (\:) =\; \inverseflow_{\tau_\theta}$.
Production is a special case of direct dialectical flow.
In particular,
if the map
$\tau_{\theta} = (\mbox{pr}_2 \cdot \theta) \oplus \tilde{d}$
associated with any marking $\theta$ over $X$ is a {\bf V}-relation,
then ordinary production by $\theta$ is direct dialectical flow along $\tau_{\theta}$;
that is,
$(\:) \oplus \theta =\; \directflow_{\tau_\theta}$.
\end{Proposition}
For the special case {\bf V} = {\bf N} when $\theta$ is finite everywhere the constraint that $\tau_{\theta}$ be a {\bf N}-relation
implies that $d$ is either $\infty$ or $0$ everywhere and that $d$ is $\infty$ iff $e$ is $\infty$.

A {\bf V}-{\em transition system} ${\sf A}$ is a triple ${\sf A} = \triple{T}{{\cal Q}}{\delta}$ consisting of:
a set (of transition symbols) $T$,
a quasi {\bf V}-space (of internal states) ${\cal Q}$, and
a transition map (transition relation assignment) $\delta \morph T \longrightarrow {\bf rel}_{\scriptbf{V}}[{\cal Q},{\cal Q}]$.
A state pair $(q,q') \memberof \product{{\cal Q}^{\rm op}}{{\cal Q}}$
is regarded as an $a$-transition from current state $q$ to next state $q'$
with weighting (probability, believability, etc.)
the generalized truth-value $\delta_a(q,q') \memberof V$ for each transition symbol $a \memberof T$.
The case ${\bf V} = {\bf R}$ of (nonnegative) real values includes probabilistic and fuzzy transition systems as special cases.
The transition map $\delta$ can be recursively extended to a ``run map'' monoid morphism
$\delta^\ast \morph T^\ast \longrightarrow {\bf rel}_{\scriptbf{V}}[{\cal Q},{\cal Q}]$:
$\delta^\ast_\varepsilon = {\rm y}_{\cal Q}$, the identity {\bf V}-relation at ${\cal Q}$; and
$\delta^\ast_{xa} = \delta^\ast_x \circ \delta_a$ for all strings of transition symbols $x \memberof T^\ast$ and single transition symbols $a \memberof T$.
Composing syntactic run map dynamics $\delta^\ast$ with flow $\yinyang$
defines the transition system dynamics
$\delta^\ast \cdot \yinyang \morph T^\ast \longrightarrow {\bf Space}_{\scriptbf{V}}[{\bf V}^{\cal Q},{\bf V}^{\cal Q}]$.

There are three main motivators for the concept of dialectical nets:
Petri nets, transition systems and Kan quantification theory of first order predicate logic.
The two motivators Petri nets and Kan quatification
correspond to the structural concepts of enriched predicates (generalized subobjects)
and enriched functions,
respectively.
The concept of dialectical nets,
which generalizes transition systems,
corresponds to the structural concept of enriched relations.

An {\em elementary dialectical {\bf V}-net}
(or {\em elementary dialectical {\bf V}-transition system})
{\sf N} is a quadruple ${\sf N} = \quadruple{T}{{\cal S}}{\iota}{o}$ consisting of:
a set (of transition symbols) $T$,
two quasi {\bf V}-spaces (of sites) ${\cal S} \!=\! \pair{{\cal S}_0}{{\cal S}_1}$,
an inverse flow assignment $\iota \morph T \longrightarrow {\bf rel}_{\scriptbf{V}}[{\cal S}_0,{\cal S}_1]$, and
a direct flow assignment $o \morph T \longrightarrow {\bf rel}_{\scriptbf{V}}[{\cal S}_0,{\cal S}_1]$,
where ${\bf rel}_{\scriptbf{V}}[{\cal S}_0,{\cal S}_1]$ is the collection
(quasi {\bf V}-space)
of all {\bf V}-relations between the quasi {\bf V}-spaces of sites ${\cal S}_0$ and ${\cal S}_1$.
So for each transition symbol $a \memberof T$ the net assigns a relation pair
$\tau_a = \parpair{{\cal S}_0}{o_a}{\iota_a}{{\cal S}_1}$ 
consisting of a direct flow specifier
${\cal S}_0 \stackrel{o_a}{\rightharpoondown} {\cal S}_1$,
and an inverse flow specifier
${\cal S}_0 \stackrel{\iota_a}{\rightharpoondown} {\cal S}_1$
which specify the dialectical flow
$\yinyang_a^{\smallsf{N}} = \yinyang_{\tau_a} = \inverseflow_{\iota_a} \cdot \directflow_{o_a}  \morph {\bf V}^{{\cal S}_1} \longrightarrow {\bf V}^{{\cal S}_1}$
for any transition symbol $a \memberof T$.
Dialectical flow $\yinyang^{\smallsf{N}}$ can be recursively extended to a ``run flow'' monoid morphism
$\yinyang^{\smallsf{N}} \morph T^\ast \longrightarrow {\bf Space}_{\scriptbf{V}}[{\bf V}^{{\cal S}_1},{\bf V}^{{\cal S}_1}]$:
$\yinyang^{\smallsf{N}}_\varepsilon = {\rm Id}_{V^{{\cal S}_1}}$, the identity {\bf V}-morphism at ${\bf V}^{{\cal P}_1}$; and
$\yinyang^{\smallsf{N}}_{xa} = \yinyang^{\smallsf{N}}_x \cdot \yinyang^{\smallsf{N}}_a$ for all strings of transition symbols $x \memberof T^\ast$ and single transition symbols $a \memberof T$.
When appropriate pullbacks exist,
the flow assignment $\tau$ can be recursively extended to a ``run map'' monoid morphism
$\tau^\ast \morph T^\ast \longrightarrow {\bf rel}_{\scriptbf{V}}[{\cal S}_1,{\cal S}_1]$.
A ``localized Beck condition'' of higher order categorical logic \cite{Lawvere70} should by incorporated into the above definition of run-flow.

The elements in the quasi {\bf V}-space ${\cal S}_1$ are the sites where values
(bit-values for conditions in condition/event nets,
numbers representing resources in consumption/production nets,
database relations in predicate/transition nets,
etc.)
are stored,
and local processing of values
(of the nature: combination, accumulation of values, suprema-calculation, union)
takes place; whereas,
the elements in the quasi {\bf V}-space ${\cal S}_0$ are usually sites where transient local processing of values
(of the dual nature: interaction, matching of values, infima-calculation, intersection)
takes place.
In the traditional theory of nets and in the traditional theory of transition systems ${\cal S}_0 = {\cal S}_1$.
In the theory of Horn clause logic programming
${\cal S}_1$ is the set of predicate (or relational) names of a logic program, and
${\cal S}_0$ is the set of clause names (or elementary implications) of same.
We regard a dialectical net to be a transformer of constrained {\bf V}-markings;
that is,
a {\bf V}-predicate transformer.
The semantics of a dialectical {\bf V}-net {\sf N} can be defined as either external or internal behaviors.
External behaviors include: (1) unfoldment-tree, and (2) regular-set behavior.
Internal behaviors include: (1) reachable predicates (markings), and (2) cumulative fixpoint behavior.
[There is a formal dialectical approach to net behavior.
Define the {\bf V}-relations
$v^\triangleleft = ((\:) \oplus v)^\triangleleft \morph {\cal V} \rightharpoondown {\cal V}$ and
$v_\triangleleft = ((\:) \oplus v)_\triangleleft \morph {\cal V} \leftharpoondown {\cal V}$.
Note that $\Box v_\triangleleft(y,x)$ iff $y \preceq x \oplus v$,
a generalized net enabling condition.
Now $v^\triangleleft$ is formally left adjoint to $v_\triangleleft$
(as an arrow in the 2-category ${\bf rel}_{\scriptbf{V}}$),
$\left( v^\triangleleft \dashv v_\triangleleft \right)$.
Let $\lhd_{\scriptbf{V}}(v)$ denote this adjunction.
Then $\lhd_{\scriptbf{V}} \morph \bullet_{\scriptbf{V}} \longrightarrow {\bf rel}_{\scriptbf{V}}$ is a formal dialectical base.
Let {\sf N} be any {\bf V}-net.
Define the dialectical flow
$\yinyang_t = (\iota_t)_\triangleleft \circ (o_t)^\triangleleft$
for each transition symbol $t \memberof T$.
For normal {\bf V},
$\Box \yinyang_t(z,x)$ iff
$z \preceq y \oplus \iota_t \mbox{ and } y \oplus o_t \preceq x \mbox{ some } y \memberof V$.
Define
$\yinyang_\varepsilon = {\rm y}_{\cal V}$;
$\yinyang_{xa} = \yinyang_x \circ \yinyang_a$ for all $x \memberof T^\ast$ and $a \memberof T$; and
$\yinyang_\ast = \bigvee_{w \in T^\ast} \yinyang_w$.
We say that $x$ is {\em reachable by} $w$ from $y$ when $\Box \yinyang_w(y,x)$; that is, when $e \preceq \yinyang_w(y,x)$.
We say that $x$ is {\em reachable} from $y$ when $\Box \yinyang_\ast(y,x)$; that is, when $e \preceq \yinyang_\ast(y,x)$.
So $\yinyang_\ast(y,x)$ gives a measure of reachability.]
Boundedness, liveness, synchronic distance and fairness
are definable for dialectical nets just as for ordinary nets.
The dialectical {\bf V}-net {\sf N} is deterministic
when $\iota$ and $o$ factor through ${\rm Space}_{\scriptbf{V}}[{\cal S}_0,{\cal S}_1]$.
Then dialectical flow is existential-quantification/substition composition
$\yinyang^{\smallsf{N}}_a = {\bf V}^{\iota_a} \cdot \exists_{o_a}$. 
Using the places-as-sites interpretation,
any {\bf V}-net ${\sf N} = \quadruple{T}{{\cal P}}{\iota}{o}$ is an elementary dialectical {\bf V}-net.
Indeed,
for the special case where
${\cal S}_0$ is the terminal-coterminal {\bf V}-space ${\cal S}_0 = {\bf 1}$
and ${\cal S}_1$ is the place-space ${\cal S}_1 = {\cal P}$
(so that ${\bf rel}_{\scriptbf{V}}[{\cal S}_0,{\cal S}_1] = {\bf rel}_{\scriptbf{V}}[{\bf 1},{\cal P}] = {\bf V}^{\cal P}$),
elementary dialectical {\bf V}-nets $\equiv$ {\bf V}-nets.
Any {\bf V}-transition system ${\sf A} = \triple{T}{{\cal Q}}{\delta}$ is an elementary dialectical {\bf V}-net.
Indeed,
for the special case where
the two site-spaces are the one state-space ${\cal S}_0 = {\cal Q} = {\cal S}_1$,
where inverse flow is the trivial identity {\bf V}-relation
$\iota_a = {\rm y}_{\cal Q}$ on ${\cal Q}$ for all transition symbols $a \memberof T$, and
where direct flow is the transition map $o = \delta$,
elementary dialectical {\bf V}-nets $\equiv$ {\bf V}-transition systems.
Using the markings-as-fuzzy-subsets interpretation,
any {\bf V}-net ${\sf N} = \quadruple{T}{P}{\iota}{o}$
is an elementary dialectical
${\bf V}^P$-net ${\sf N} = \quadruple{T}{{\bf 1}}{\iota}{o}$ 
with only one site
${\cal S}_0 = {\bf 1} = {\cal S}_1$
(so that ${\bf rel}_{V^P}[{\cal S}_0,{\cal S}_1] = {\bf rel}_{V^P}[{\bf 1},{\bf 1}] \cong {\bf V}^P$),
where inverse flow 
$\iota \morph T \longrightarrow {\bf V}^P \cong {\bf rel}_{V^P}[{\bf 1},{\bf 1}]$ and
direct flow $o \morph T \longrightarrow {\bf V}^P \cong {\bf rel}_{V^P}[{\bf 1},{\bf 1}]$
assign ${\bf V}^P$-elements as ${\bf V}^P$-relations.
That is,
elementary dialectical ${\bf V}^P$-nets on one site $\equiv$ {\bf V}-nets.
Here $\yinyang_t(\mu) = [\iota_t \!\Rightarrow\! \mu] \oplus o_t = {\sf N}_t(\mu)$
for any transition symbol $t \memberof T$;
that is,
dialectical flow is ordinary consumption/production transitioning.
These are not transition systems since inverse flow is not identity.
 
Any two {\bf V}-relations in the collection
$\{ {\cal S}_0 \stackrel{\iota_a}{\rightharpoondown} {\cal S}_1 \mid a \memberof T \}$
of inverse flow assignments
are obviously comparable by use of the inf metric on the quasi {\bf V}-space
${\bf V}^{\subsupsize{\product{{\cal S}_0^{\rm op}}{{\cal S}_1}}}$.
We can intend, or specify, relationships between the $\iota_a$ by requiring
that $T$ be a general (and not just discrete) quasi {\bf V}-space ${\cal T} = \pair{T}{d_T}$, and
that inverse flow assignment be a {\bf V}-morphism
$\iota \morph {\cal T}^{\rm op} \longrightarrow {\bf rel}_{\scriptbf{V}}[{\cal S}_0,{\cal S}_1] \!=\! {\bf V}^{\subsupsize{\product{{\cal S}_0^{\rm op}}{{\cal S}_1}}}$. 
The same comments apply to the direct flow assignment $o$.
A {\em dialectical {\bf V}-net}
(or {\em dialectical {\bf V}-transition system})
{\sf N} is a quadruple ${\sf N} = \quadruple{{\cal T}}{{\cal S}}{\iota}{o}$ consisting of:
a quasi {\bf V}-space of transition symbols ${\cal T}$,
two quasi {\bf V}-spaces of sites ${\cal S} \!=\! \pair{{\cal S}_0}{{\cal S}_1}$,
an inverse flow {\bf V}-morphism $\iota \morph {\cal T}^{\rm op} \longrightarrow {\bf rel}_{\scriptbf{V}}[{\cal S}_0,{\cal S}_1]$, and
a direct flow {\bf V}-morphism $o \morph {\cal T}^{\rm op} \longrightarrow {\bf rel}_{\scriptbf{V}}[{\cal S}_0,{\cal S}_1]$.
Dialectical nets have the same run-flow dynamics as elementary dialectical nets.
It is clear that every dialectical net
${\sf N} = \quadruple{{\cal T}}{{\cal S}}{\iota}{o}$
defines a (single) dialectical flow specifier.
Both the input and the output weighting functions are {\bf V}-morphisms
$\iota,o \morph \product{\product{{\cal T}^{\rm op}}{{\cal S}_0^{\rm op}}}{{\cal S}_1} \longrightarrow {\bf V}$;
that is,
{\bf V}-relations
$\iota,o \morph \product{{\cal T}}{{\cal S}_0} \rightharpoondown {\cal S}_1$.
This means that the dialectical net {\sf N} is just the (single) relation pair
${\sf N} = \;\bigparpair{\product{{\cal T}}{{\cal S}_0}}{o}{\iota}{{\cal S}_1}$,
which can be viewed as an enriched {\em state-space graph} of {\sf N}.
For a {\bf V}-transition system ${\sf A} = \triple{T}{{\cal Q}}{\delta}$,
$\product{{\cal T}}{{\cal S}_0} = \product{T}{{\cal Q}}$
is the $T$-th copower of ${\cal Q}$.
If the transition system {\sf A} is discrete and deterministic,
$o = \delta \morph \product{T}{Q} \longrightarrow Q$ is the usual determistic state transition function,
$\iota = {\rm pr}_Q \morph \product{T}{Q} \longrightarrow Q$ is the projection function (identity inverse flow),
$\product{T}{Q}$ is the set of edges in the state-space,
and the relation pair
${\sf A} = \;\bigparpair{\product{{\cal T}}{{\cal Q}}}{\delta}{{\rm pr}_Q}{{\cal Q}}$
is the ordinary state-space graph for the transition system {\sf A}.
If we interpret the state-space graph of a dialectical net {\sf N} to be a dialectical flow specifier
we can define an aggregate dialectical flow
$\yinyang^{\smallsf{N}} \morph {\bf V}^{{\cal S}_1} \longrightarrow {\bf V}^{{\cal S}_1}$
on {\bf V}-predicates over the site-space ${\cal S}_1$.
This aggregate dialectical flow,
and its various fixpoints,
define a combined external-internal behavior for the dialectical net {\sf N}.
For a fixed space of transition symbols ${\cal T}$,
given any two dialectical {\bf V}-nets
$\pair{{\cal S}_0}{\tau_0} = \triple{{\cal S}_0}{\iota_0}{o_0}$ and
$\pair{{\cal S}_1}{\tau_1} = \triple{{\cal S}_1}{\iota_1}{o_1}$,
a {\em morphism} of dialectical {\bf V}-nets
$\pair{h_0}{h_1} \morph \triple{{\cal S}_0}{\iota_0}{o_0} \Rightarrow \triple{{\cal S}_1}{\iota_1}{o_1}$
consists of two {\bf V}-morphisms
$h_0 \morph {\cal S}_{0,0} \longrightarrow {\cal S}_{1,0}$ and
$h_1 \morph {\cal S}_{0,1} \longrightarrow {\cal S}_{1,1}$,
where
$\pair{h_0}{h_1} \morph \pair{{\cal S}_0}{\tau_{0,a}} \Rightarrow \pair{{\cal S}_1}{\tau_{1,a}}$
is a vertical morphism in ${\bf Rel}_{\scriptbf{V}}^\times$ for all transition symbols $a \memberof T$.
For fixed ${\cal T}$,
dialectical {\bf V}-nets and their morphisms form a category
${\bf Net}_{\scriptbf{V}}^{\cal T}$,
the ${\cal T}$-th fiber of the
{\em category of dialectical {\bf V}-nets}
${\bf Net}_{\scriptbf{V}}$.

{\bf V}-predicates and flow conditions form a category ${\bf Pred}_{\scriptbf{V}}$,
whose objects are pairs $\pair{{\cal X}}{\phi}$
where ${\cal X}$ is a quasi {\bf V}-space
and $\phi \memberof {\bf V}^{{\cal X}}$ is a {\bf V}-predicate over ${\cal X}$,
and whose morphisms
$\pair{{\cal Y}}{\psi} \stackrel{\tau}{\longrightarrow} \pair{{\cal X}}{\phi}$
are {\bf V}-relations
${\cal Y} \stackrel{\tau}{\rightharpoondown} {\cal X}$
satisfying the direct flow condition
$\directflow_\tau\!(\psi) \preceq_{\cal X} \phi$
or the equivalent inverse flow condition
$\psi \preceq_{\cal Y}\; \inverseflow_\tau\!(\phi)$
There is a underlying {\bf V}-space functor
$P_{\scriptbf{V}} \morph {\bf Pred}_{\scriptbf{V}} \longrightarrow {\bf rel}_{\scriptbf{V}}$ 
where $P_{\scriptbf{V}}(\pair{{\cal X}}{\phi}) = {\cal X}$
and $P_{\scriptbf{V}}(\tau) = \tau$.
A {\em dialectical base} is a 01-fibration
$P \morph {\bf E} \longrightarrow {\bf \Omega}$
whose fibers ${\bf E}_w = P^{-1}(w)$ are bicomplete.
\begin{Proposition}
{\bf V}-predicates and flow conditions form a dialectical base over
{\bf V}-spaces and {\bf V}-relations;
that is,
the functor
$P_{\scriptbf{V}} \morph {\bf Pred}_{\scriptbf{V}} \longrightarrow {\bf rel}_{\scriptbf{V}}$
is a dialectical base.
\end{Proposition}
This dialectical base erects a second external level of dialectical structure
over the first internal level of dialectical structure represented by the category ${\bf rel}_{\scriptbf{V}}$.
It will prove useful in the definition of the type theory of {\bf V}-nets,
and in the recursive specification of {\bf V}-nets.
Recall that in order to motivate the notion of a quasi {\bf V}-space,
we showed how to translate external marking constraints
such as net transition enabling conditions or the more general form $\mu_1 \preceq \mu_0$
into internal marking constraints (or metrics) $d$.
However,
with the introduction of the category of {\bf V}-predicates ${\bf Pred}_{\scriptbf{V}}$
we have incorporated both internal and external constraints into our dialectical approach.
The internal constraints on markings $\phi \morph X \rightarrow {\bf V}$
are still specified by metrics $d \morph \product{X}{X} \rightarrow {\bf V}$ on $X$,
and markings which satisfy internal constraints $d$ are called {\bf V}-predicates $\phi \memberof {\bf V}^{{\cal X}}$ over ${\cal X} \!=\! \pair{X}{d}$.
So internal constraints are embedded into the objects of ${\bf Pred}_{\scriptbf{V}}$.
We identify external constraints with the morphisms of ${\bf Pred}_{\scriptbf{V}}$:
a morphism $\pair{{\cal X}}{\phi} \stackrel{\tau}{\rightharpoondown} \pair{{\cal Y}}{\psi}$
imposes the external dialectical constraint
$\directflow_\tau(\psi) \preceq_{\cal X} \phi$ or equivalently
$\psi \preceq_{\cal Y} \inverseflow_\tau(\phi)$
on markings (now called {\bf V}-predicates).
These external constraints include the original external constraints.
In fact the original external constraints are specified precisely by the identities:
the identity {\bf V}-relation
$\tau = d_X = \mbox{Id}_{\cal X}$
on ${\cal X}$ gives the external dialectical constraint
$\psi \preceq_{\cal X} \phi$.
But, in general, the external constraints are no longer necessarily pointwise constraints.

In our interpretation of dialectical flow
the quasi {\bf V}-space ${\cal S}_0$ is a site of transient entities (predicates),
whereas the quasi {\bf V}-space ${\cal S}_1$ is the site where predicates actually reside.
So we can allow the transient site-space ${\cal S}_0$ to vary as transient symbols $a \memberof T$ vary.
A {\em generalized dialectical {\bf V}-net} {\sf N} is a pair
${\sf N} = \pair{{\bf T}}{\tau}$ consisting of:
a graph (or category) of transition symbols ${\bf T}$, and
a cocone $\tau \morph S_0 \Longrightarrow {\cal S}_1$ of relation pairs
whose base is a diagram (or functor) $S_0 \morph {\bf T}^{\rm op} \longrightarrow {\bf Flow}_{\scriptbf{V}}$ of transient site-spaces
and whose vertex is a quasi {\bf V}-space of sites ${\cal S}_1$.
Let $\tilde{{\cal S}_0} = {\rm Colim}(S_0) = \prod_{a \in |\scriptbf{T}|}S_{0,a}$
be the colimit of $S_0$ in ${\bf Flow}_{\scriptbf{V}}$.
Then any generalized dialectical net {\sf N} is equivalent to the unique flow specifier
$\tilde{\tau} = \parpair{\tilde{{\cal S}_0}}{o}{\iota}{{\cal S}_1}$ determined by the cocone $\tau$.
If $S_0$ is constant on objects (transition symbols),
$S_{0,a} = {\cal S}_0$ for all $a \memberof |{\bf T}|$,
then $\tilde{{\cal S}_0} = \product{|{\bf T}|}{{\cal S}_0}$. \doublespace


\hspace*{\fill} DIALECTICAL SYSTEMS \hspace*{\fill} \doublespace
In this section we show that each dialectical net determines a special kind of dialectical system.
Let
${\bf V} = \quintuple{V}{\preceq}{\oplus}{\Rightarrow}{e}$
be any closed preorder.
The monoid $\triple{V}{\oplus}{e}$ can be regarded to be a category,
denoted by $\bullet_V$.
Let us give the monotonic functions of
{\bf V}-composition and {\bf V}-implication
the more function-like notation
$V^v(\:) = (\:) \oplus v \morph V \longrightarrow V$ and
$V_v(\:) = v \!\Rightarrow\! (\:)  \morph V \longrightarrow V$
for each {\bf V}-element $v \memberof V$.
Then the closure axiom becomes the adjunction statement
$(V^v \dashv V_v) \morph V \longrightarrow V$
for each {\bf V}-element $v \memberof V$.
Let ${\bf V}(n)$ denote this adjunction. 
In objective dialectics,
since dialectical contradictions are represented by adjunctions,
systems of dialectical contradictions are represented by diagrams (pseudofunctors) in the category {\bf adj}
whose objects are bicomplete preorders and whose morphisms are adjoint pairs of monotonic functions. 
We call such a (pseudo)functor ${\bf E} \morph {\bf \Omega} \longrightarrow {\bf adj}$
a {\em dialectical base} of preorders,
and use the notation
${\bf E}(w_1 \stackrel{t}{\rightarrow} w_2) = ({\bf E}^t \dashv {\bf E}_t) \morph {\bf E}_{w_1} \rightarrow {\bf E}_{w_2}$.
Objects of ${\bf \Omega}$ are called {\em types} and arrows of ${\bf \Omega}$ are called {\em terms}.
Dialectical systems are the ``motors of nature'' specifying the dialectical motion of structured entities,
and a dialectical base provides the ``motive power'' for this motion.
The entire set of axioms for the closed preorder
${\bf V} = \quintuple{V}{\preceq}{\oplus}{\Rightarrow}{e}$
are equivalent to the following single statement.
\begin{Fact}
The operator ${\bf V}(\:)$ is a dialectical base
${\bf V} \morph \bullet_V \longrightarrow {\bf adj}$.
\end{Fact}

We now develop an equivalent fibrational approach for formalizing the dialectical structure of the {\bf V}-elements.
Any quasi {\bf V}-space ${\cal X} = \pair{X}{d_X}$ determines a category ${\bf X}$,
whose objectset is ${\rm Obj}({\bf X}) = X$,
whose arrowset is ${\rm Ar}({\bf X}) = \{(x,v,x') \mid v \preceq d_X(x,x')\}$ with homsets being the principal ideals ${\bf X}[x,x'] \cong\; \downarrow_V\!\!d_X(x,x')$,
whose source and target functions are the projections ${\rm pr}_1,{\rm pr}_3 \morph {\rm Ar}({\bf X}) \longrightarrow X$,
whose identities are ${\rm Id}_x = (x,e,x)$ for each $x \memberof X$, and
whose composition is $(x,v,x') \circ (x',v',x'') = (x,v \oplus v',x'')$.
Clearly the projection function ${\rm pr}_2 \morph {\rm Ar}({\bf X}) \longrightarrow V$
defines a functor $|\:|_{\cal X} = {\rm pr}_2 \morph {\bf X} \longrightarrow \bullet_V$.
Since $(x,e,x')$ is an ${\bf X}$-arrow
iff $e \preceq d_X(x,x')$
iff $x \preceq_X x'$,
the fiber of $|\:|_{\cal X}$ over the (only) $\bullet_V$-object $e$ is
$|\:|_{{\cal X},e} = |\:|_{\cal X}^{-1}(e) = \{(x,e,x') \mid e \preceq d_X(x,x')\}
\cong \{(x,x') \mid x \preceq_X x'\}$;
that is,
the fiber $|\:|_{{\cal X},e}$ is essentially the preorder $\pair{X}{\preceq_X}$ viewed as a subcategory of {\bf X}.
A {\em distributed} {\bf V}-{\em type} (or a {\bf V}-{\em normed category}) ${\cal C}$ is a pair ${\cal C} = \pair{{\bf C}}{|\:|_{\cal C}}$
where ${\bf C}$ is a category and $|\:|_{\cal C} \morph {\bf C} \longrightarrow \bullet_V$ is a functor called a {\bf V}-{\em norm} (or a {\bf V}-{\em typing}).
So every quasi {\bf V}-space ${\cal X} = \pair{X}{d_X}$ determines a distributed {\bf V}-type ${\cal X} = \pair{{\bf X}}{|\:|_{\cal X}}$.
Any category {\bf C} is a distributed {\bf V}-type ${\cal C} = \pair{{\bf C}}{e}$,
where $e \morph {\bf C} \rightarrow \bullet_V$ is the constant identity functor.
The category {\bf V} of {\bf V}-{\em inequality conditions} is the category part of
$\pair{{\bf V}}{|\:|_{\cal V}}$,
the distributed {\bf V}-type determined by the quasi {\bf V}-space ${\cal V}$ of generalized truth-values.
The functor part $|\:|_{\cal V}$ has special properties, and so we give it a special notation:
$P_V = |\:|_{\cal V}$.
By the above fact, $P_V$ is a $01$-fibration.
The only fiber of $P_V$ is $P_{V,e} \cong {\cal V}=\pair{V}{\preceq}$ which is bicomplete (a cpo).
\begin{Proposition}
{\bf V}-inequality conditions form a dialectical base over {\bf V}-truth-values;
that is,
the {\bf V}-norm
$P_V \morph {\bf V} \longrightarrow \bullet_V$ is a dialectical base.
\end{Proposition}

We can extend this result from ${\bf V}=\triple{{\cal V}}{\oplus}{\Rightarrow}$ itself to any tensored-cotensored quasi {\bf V}-space .
Any {\bf V}-morphism ${\cal X} \stackrel{f}{\longrightarrow} {\cal Y}$ of quasi {\bf V}-spaces ${\cal X} = \pair{X}{d_X}$ and ${\cal Y} = \pair{Y}{d_Y}$
determines a functor ${\bf X} \stackrel{H_f}{\longrightarrow} {\bf Y}$
where $H_f(x) = f(x)$ on {\bf X}-objects
and $H_f((x,v,x')) = (f(x),v,f(x'))$ on {\bf X}-arrows.
Clearly $H_f$ commutes with the projection functions:
$H_f \cdot |\:|_{\cal Y} = |\:|_{\cal X}$.
Let ${\rm Type}_V(f) = H_f$ denote this construction.
A {\em morphism} of distributed {\bf V}-types
$H \morph \pair{{\bf C}}{|\:|_{\cal C}} \longrightarrow \pair{{\bf D}}{|\:|_{\cal D}}$
is a functor $H \morph {\bf C} \longrightarrow {\bf D}$
which commutes with the {\bf V}-norms:
$H \cdot |\:|_{\cal C} = |\:|_{\cal D}$.
So every quasi {\bf V}-space morphism is a distributed {\bf V}-type morphism.
A {\bf V}-predicate $\phi \memberof {\bf V}^{\cal X}$ over a quasi {\bf V}-space ${\cal X}$
is a {\bf V}-morphism $\phi \morph {\cal X} \longrightarrow {\bf V}$,
and hence determines a morphism of distributed {\bf V}-types
$H_\phi \morph {\rm Type}_V({\cal X}) \longrightarrow {\rm Type}_V({\cal V})$;
that is,
a functor $H_\phi \morph {\bf X} \longrightarrow {\bf V}$ satisfying $H_\phi \cdot P_V = |\:|_{\cal X}$.
Given a distributed {\bf V}-type ${\cal C} = \pair{{\bf C}}{|\:|_{\cal C}}$,
a {\em distributed} {\bf V}-{\em entity} (or {\bf V}-{\em predicate}) $\Phi$ of type ${\cal C}$
is a functor $\Phi \morph {\bf C} \longrightarrow {\bf V}$ satisfying $\Phi \cdot P_V = |\:|_{\cal C}$.
Let ${\bf V}^{\cal C}$ denote the collection (bicomplete quasi {\bf V}-space) of all distributed {\bf V}-entities of type ${\cal C}$.
Let ${\bf Type}_V$ denote the category of distributed {\bf V}-types and their morphisms.
${\bf Type}_V$ is the comma category ${\bf Type}_V = {\bf Cat}\!\downarrow\!\bullet_V$.
Then ${\rm Type}_V$ is a functor
${\rm Type}_V \morph {\bf Space}_V \longrightarrow {\bf Type}_V$
from spaces to types.

In the opposite direction,
any distributed {\bf V}-type ${\cal C} = \pair{{\bf C}}{|\:|_{\cal C}}$ determines a quasi {\bf V}-space ${\cal C} = \pair{C}{d_C}$,
where $C$ is the objectset $C = {\rm Obj}({\bf C})$ of {\bf C}
and the metric $d_C \morph \product{C}{C} \longrightarrow V$ is defined to be the homset supremum
$d_C(c,c') = \bigvee_{c \stackrel{g}{\rightarrow} c'}|g|_{\cal C}$.
Let ${\rm Space}_V(\pair{{\bf C}}{|\:|_{\cal C}}) = \pair{C}{d_C}$ denote this construction.
Any morphism of distributed {\bf V}-types $H \morph {\cal C} \longrightarrow {\cal D}$,
where ${\cal C} = \pair{{\bf C}}{|\:|_{\cal C}}$ and ${\cal D} = \pair{{\bf D}}{|\:|_{\cal D}}$,
determines a morphism of quasi {\bf V}-spaces $f_H \morph {\rm Space}_V({\cal C}) \longrightarrow {\rm Space}_V({\cal D})$,
where $f_H(c) = H(c)$ for all objects $c \memberof {\bf C}$.
Let ${\rm Space}_V(H) = f_H$ denote this construction.
Then ${\rm Space}_V$ is a functor
${\rm Space}_V \morph {\bf Type}_V \longrightarrow {\bf Space}_V$
from types to spaces.
For any distributed {\bf V}-type ${\cal C} = \pair{{\bf C}}{|\:|_{\cal C}}$
there is a canonical morphism of {\bf V}-types
$\eta_{{\cal C}} \morph {\cal C} \longrightarrow {\rm Type}_V({\rm Space}_V({\cal C}))$,
where $\eta_{\cal C}$ is the identity map on objects
and $\eta_{\cal C}(c \stackrel{g}{\rightarrow} c') = c \stackrel{(c,|g|_{\cal C},c')}{\longrightarrow} c'$ on arrows.
Moreover, ${\rm Space}_V({\rm Type}_V({\cal X})) = {\cal X}$ for any quasi {\bf V}-space ${\cal X}$.
\begin{Proposition}
The {\bf V}-space functor is left adjoint $({\rm Space}_V \dashv {\rm Type}_V)$ to the {\bf V}-type functor.
\end{Proposition}
The unit of this adjunction $\eta \morph {\rm Id} \Longrightarrow {\rm Space}_V \cdot {\rm Type}_V$
has the canonical morphism of {\bf V}-types $\eta_{\cal C}$ as its ${\cal C}$-th component,
and the counit of this adjunction is the identity natural transformation ${\rm Id} \morph {\rm Type}_V \cdot {\rm Space}_V \Longrightarrow {\rm Id}$.
So this adjunction is a reflection with
${\rm Type}_V \morph {\bf Space}_V \longrightarrow {\bf Type}_V$
embedding ${\bf Space}_V$ as a subcategory of ${\bf Type}_V$, and
${\rm Space}_V \morph {\bf Type}_V \longrightarrow {\bf Space}_V$
reflecting ${\bf Type}_V$ into its ``subcategory'' ${\bf Space}_V$.

Given two categories {\bf B} and {\bf A},
a {\em distributor} ${\bf R}$ from category {\bf B} to category {\bf A},
denoted by ${\bf B} \stackrel{\scriptbf{R}}{\rightharpoondown} {\bf A}$,
is a triple ${\bf R} = \triple{\circ_0}{R}{\circ_1}$,
where:
$R$ is a span $R = \quintuple{{\rm Obj}({\bf B})}{\partial_0}{{\rm Ar}({\bf R})}{\partial_1}{{\rm Obj}({\bf A})}$
with arrowset ${\rm Ar}({\bf R})$, source function $\partial_0 \morph {\rm Ar}({\bf R}) \longrightarrow {\rm Obj}({\bf B})$,
and target function $\partial_1 \morph {\rm Ar}({\bf R}) \longrightarrow {\rm Obj}({\bf A})$;
$\circ_0$ is a left action with respect to {\bf B}, so that
${\rm Id}_b \circ_0 e = e$ and $(g' \circ_B g) \circ_0 e = g' \circ_0 (g \circ_0 e)$
for all $R$-arrows $b \stackrel{e}{\rightarrow} a$ and all {\bf B}-arrows $b'' \stackrel{g'}{\rightarrow} b'$ and $b' \stackrel{g}{\rightarrow} b$;
$\circ_1$ is a right action with respect to {\bf A}, so that
$e \circ_1 {\rm Id}_a = e$ and $e \circ_1 (f \circ_A f') = (e \circ_1 f) \circ_1 f'$
for all $R$-arrows $b \stackrel{e}{\rightarrow} a$ and all {\bf A}-arrows $a \stackrel{f}{\rightarrow} a'$ and $a' \stackrel{f'}{\rightarrow} a''$;
and the mixed associative law
$g \circ_0 (e \circ_1 f) = (g \circ_0 e) \circ_1 f)$
holds.
A category ${\bf C}$ is a distributor
${\bf C} \stackrel{\scriptbf{C}}{\rightharpoondown} {\bf C}$,
where ${\bf C} = \triple{\circ_C}{C}{\circ_C}$
and $C = \quintuple{{\rm Obj}({\bf C})}{\partial_0^C}{{\rm Ar}({\bf C})}{\partial_1^C}{{\rm Obj}({\bf C})}$.
Categories and distributors form a category {\bf dist},
which includes as a subcategory (via Yoneda)
the category {\bf Cat} of categories and functors.
A {\em morphism of distributors}
${\bf R}_1 \stackrel{\scriptbf{H}}{\Longrightarrow} {\bf R}_2$
from distributor ${\bf B}_1 \stackrel{\scriptbf{R}_1}{\rightharpoondown} {\bf A}_1$
to distributor ${\bf B}_2 \stackrel{\scriptbf{R}_2}{\rightharpoondown} {\bf A}_2$,
is a triple 
${\bf H} = \triple{G}{H}{F}$
where $G \morph {\bf B}_1 \longrightarrow {\bf B}_2$
and $F \morph {\bf A}_1 \longrightarrow {\bf A}_2$
are functors,
and $\triple{G}{H}{F} \morph R_1 \longrightarrow R_2$ is a morphism of spans
which preserves actions:
$H$ preserves source, $\partial_0(H(e)) = G(\partial_0(e))$;
$H$ preserves target, $\partial_1(H(e)) = F(\partial_1(e))$; 
$H$ preserves source (left) action, $H(g \circ_0 e) = G(g) \circ_0 H(e)$; and
$H$ preserves target (right) action, $H(e \circ_1 f) = H(e) \circ_1 F(f)$;
for all
$b' \stackrel{g}{\rightarrow} b$ in ${\rm Ar}({\bf B})$,
$b \stackrel{e}{\rightarrow} a$ in ${\rm Ar}({\bf R})$ and
$a \stackrel{f}{\rightarrow} a'$ in ${\rm Ar}({\bf A})$.
Distributers and their vertical morphisms form the category {\bf Dist}.
A functor ${\bf A}_1 \stackrel{F}{\longrightarrow} {\bf A}_2$ is a morphism of distributors,
${\bf A}_1 \stackrel{\scriptbf{F}}{\Longrightarrow} {\bf A}_2$
with categories ${\bf A}_1$ and ${\bf A}_2$ regarded as distributors,
where ${\bf F} = \triple{F \morph {\rm Obj}({\bf A}_1) \rightarrow {\rm Obj}({\bf A}_2)}
                        {F \morph {\rm Ar}({\bf A}_1) \rightarrow {\rm Ar}({\bf A}_2)}
                        {F \morph {\rm Obj}({\bf A}_1) \rightarrow {\rm Obj}({\bf A}_2)}$. 
This defines a vertical embedding of {\bf Cat} into {\bf Dist}.

Any {\bf V}-relation ${\cal Y} \stackrel{\tau}{\rightharpoondown} {\cal X}$
determines a distributor ${\bf Y} \stackrel{\scriptbf{T}}{\rightharpoondown} {\bf X}$
whose arrowset is ${\rm Ar}({\bf T}) = \{(y,v,x) \mid y \memberof Y, x \memberof X, v \preceq \tau(y,x)\}$
with homsets being the principal ideals ${\bf T}[y,x] \cong\: \downarrow_V\! \tau(y,x)$,
whose source and target functions are the projections
$\partial_0^T = {\rm pr}_1 \morph {\rm Ar}({\bf T}) \longrightarrow {\rm Obj}({\bf Y})$ and
$\partial_1^T = {\rm pr}_3 \morph {\rm Ar}({\bf T}) \longrightarrow {\rm Obj}({\bf X})$,
and whose biaction consists of
the left action $(y',v,y) \circ_0 (y,w,x) = (y',v \oplus w,x)$ and
the right action $(y,w,x) \circ_1 (x,u,x') = (y,w \oplus u,x')$
(essentially the {\bf V}-composition $\oplus$ distributed over $Y$ and $X$).
Clearly the projection function ${\rm pr}_2 \morph {\rm Ar}({\bf T}) \longrightarrow V$
defines a morphism of spans $|\:|_{\cal T} = {\rm pr}_2 \morph {\bf T} \longrightarrow \bullet_V$
which preserves left and right actions:
$|(y',v,y) \circ_0 (y,w,x)|_{\cal T} = v \oplus w = |(y',v,y)|_{\cal Y} \oplus |(y,w,x)|_{\cal T}$.
So ${\bf T} \stackrel{|\:|_{\cal T}}{\Longrightarrow} \bullet_V$ is a morphism of distributors,
where $|\:|_{\cal T} = \triple{|\:|_{\cal Y}}{|\:|_{\cal T}}{|\:|_{\cal X}}$.
Since $(y,e,x)$ is an ${\bf T}$-arrow
iff $e \preceq \tau(y,x)$
iff $y \preceq_\tau x$,
the fiber of $|\:|_{\cal T}$ over the (only) $\bullet_V$-object $e$ is
$|\:|_{{\cal T},e} = |\:|_{\cal T}^{-1}(e) = \{(y,e,x) \mid e \preceq \tau(y,x)\}
\cong \{(y,x) \mid y \preceq_\tau x\}$;
that is,
the fiber $|\:|_{{\cal X},e}$ is essentially the {\bf 2}-relation ${\cal Y} \stackrel{\preceq_\tau}{\rightharpoondown} {\cal X}$
viewed as a subdistributor of ${\bf Y} \stackrel{\scriptbf{T}}{\rightharpoondown} {\bf X}$.
Given two distributed types ${\cal B}$ and ${\cal A}$,
a {\em distributed} {\bf V}-{\em term} (or a {\bf V}-{\em normed distributor})
${\cal R}$ from distributed {\bf V}-type ${\cal B}$ to distributed {\bf V}-type ${\cal A}$,
denoted by ${\cal B} \stackrel{\scriptcal{R}}{\rightharpoondown} {\cal A}$,
is a pair ${\cal R} = \pair{{\bf R}}{|\:|_{\cal R}}$
where ${\bf B} \stackrel{\scriptbf{R}}{\rightharpoondown} {\bf A}$ is a distributor
and ${\bf R} \stackrel{|\:|_{\cal R}}{\Longrightarrow} \bullet_V$ is a morphism of distributors,
called a {\bf V}-{\em norm} (or a {\bf V}-{\em terming}),
where $|\:|_{\cal R} = \triple{|\:|_{\cal B}}{|\:|_{\cal R}}{|\:|_{\cal A}}$.
So every {\bf V}-relation ${\cal Y} \stackrel{\tau}{\rightharpoondown} {\cal X}$ determines a distributed {\bf V}-term ${\cal T} = \pair{{\bf T}}{|\:|_{\cal T}}$.
Let ${\rm Term}_V(\tau) = \pair{{\bf T}}{|\:|_{\cal T}}$ denote this construction.
Also, a distributed {\bf V}-type ${\cal C}$ is a distributed {\bf V}-term
${\cal C} \stackrel{{\cal C}}{\rightharpoondown} {\cal C}$ where ${\cal C} = \pair{{\bf C}}{|\:|_{\cal C}}$.
Any distributor
${\bf B} \stackrel{\scriptbf{R}}{\rightharpoondown} {\bf A}$
is a distributed {\bf V}-term
${\cal B} \stackrel{\scriptcal{R}}{\rightharpoondown} {\cal A}$
with constant identity {\bf V}-norm
${\bf R} \stackrel{e}{\Longrightarrow} \bullet_V$.
So {\bf Dist} is embeddable into ${\bf Term}_V$.
So distributed {\bf V}-types and distributed {\bf V}-terms form a category ${\bf term}_V$.
There is a concept orthogonal to distributed-terms-as-morphisms.
Any morphism $\tau_1 \stackrel{\mbox{\footnotesize$\pair{g}{f}$\normalsize}}{\longrightarrow} \tau_2$ of {\bf V}-relations ${\cal Y}_1 \stackrel{\tau_1}{\rightharpoondown} {\cal X}_1$ and ${\cal Y}_2 \stackrel{\tau_2}{\rightharpoondown} {\cal X}_2$
determines a morphism of distributors ${\bf T}_1 \stackrel{\scriptbf{H}_{g,f}}{\longrightarrow} {\bf T}_2$
where ${\bf H}_{g,f} = \triple{|\:|_{\cal Y}}{H_{g,f}}{|\:|_{\cal X}}$
with $H_{g,f}((y,v,x)) = (g(y),v,f(x))$ on ${\bf T}_1$-arrows.
Clearly ${\bf H}_{g,f}$ commutes with the {\bf V}-norms:
${\bf H}_{g,f} \cdot |\:|_{{\cal T}_2} = |\:|_{{\cal T}_1}$.
Let
${\rm Term}_V({\bf A}) = {\rm Space}({\bf A})$ and
${\rm Term}_V(\pair{g}{f}) = {\bf H}_{g,f}$
denote this construction.
Given two distributed {\bf V}-terms
${\cal R}_1 = \pair{{\bf R}_1}{|\:|_{{\cal R}_1}}$ and  ${\cal R}_2 = \pair{{\bf R}_2}{|\:|_{{\cal R}_2}}$,
a {\em morphism of distributed {\bf V}-terms}
${\cal R}_1 \stackrel{\scriptbf{H}}{\Longrightarrow} {\cal R}_2$
is a morphism of distributors
${\bf R}_1 \stackrel{\scriptbf{H}}{\Longrightarrow} {\bf R}_2$,
say ${\bf H} = \triple{G}{H}{F}$,
which commutes with the {\bf V}-norms:
${\bf H} \cdot |\:|_{{\cal R}_1} = |\:|_{{\cal R}_2}$.
So every morphism of {\bf V}-relations is a morphism of distributed {\bf V}-terms.
Let ${\bf Term}_V$ denote the category of distributed {\bf V}-terms and their morphisms.
${\bf Term}_V$ is the comma category
${\bf Term}_V = {\bf Dist}\!\Downarrow\!\bullet_V$.
This is the vertical category of a double category,
which we also denote by ${\bf Term}_V$,
whose underlying horizontal category is ${\bf term}_V$.
Then ${\rm Term}_V$ is a functor
${\rm Term}_V \morph {\bf Rel}_V \longrightarrow {\bf Term}_V$
from relations to terms.


In the opposite direction,
any distributed {\bf V}-term ${\cal T} = \pair{({\bf B} \stackrel{\scriptbf{T}}{\rightharpoondown} {\bf A})}{|\:|_{\cal T}}$ determines a {\bf V}-relation ${\rm Space}({\bf B}) \stackrel{\tau_T}{\rightharpoondown} {\rm Space}({\bf A})$,
where the {\bf V}-morphism $\tau_T \morph \product{B}{A} \longrightarrow V$ is defined to be the homset supremum
$\tau_T(b,a) = \bigvee_{b \stackrel{e}{\rightarrow} a}|e|_{\cal T}$.
Let ${\rm Rel}_V(\pair{{\bf T}}{|\:|_{\cal T}}) = \tau_T$ denote this construction.
Any morphism of distributed {\bf V}-terms $F\!=\!\triple{g}{h}{f} \morph {\cal T}_1 \longrightarrow {\cal T}_2$,
where ${\cal T}_1 = \pair{{\bf T}_1}{|\:|_{{\cal T}_1}}$ and ${\cal T}_2 = \pair{{\bf T}_2}{|\:|_{{\cal T}_2}}$,
determines a morphism of {\bf V}-relations $\pair{g}{f} \morph {\rm Rel}_V({\cal T}_1) \longrightarrow {\rm Rel}_V({\cal T}_2)$.
Let ${\rm Rel}_V(F) = \pair{g}{f}$ denote this construction.
Then ${\rm Rel}_V$ is a functor
${\rm Rel}_V \morph {\bf Term}_V \longrightarrow {\bf Rel}_V$
from terms to relations.
For any distributed {\bf V}-term ${\cal T} = \pair{{\bf T}}{|\:|_{\cal T}}$
there is a canonical morphism of {\bf V}-terms
$\eta_{{\cal T}} \morph {\cal T} \longrightarrow {\rm Term}_V({\rm Rel}_V({\cal T}))$,
where $\eta_{\cal T} = \triple{{\rm Id}_B}{h}{{\rm Id}_A}$
and $h(b \stackrel{e}{\rightarrow} a) = b \stackrel{(b,|e|_{\cal T},a)}{\longrightarrow} a$ on arrows.
Moreover, ${\rm Rel}_V({\rm Term}_V(\tau)) = \tau$ for any {\bf V}-relation $\tau$.
\begin{Proposition}
The {\bf V}-relation functor is left adjoint $({\rm Rel}_V \dashv {\rm Term}_V)$ to the {\bf V}-term functor.
\end{Proposition}
The unit of this adjunction $\eta \morph {\rm Id} \Longrightarrow {\rm Rel}_V \cdot {\rm Term}_V$
has the canonical morphism of {\bf V}-terms $\eta_{\cal T}$ as its ${\cal T}$-th component,
and the counit of this adjunction is the identity natural transformation ${\rm Id} \morph {\rm Term}_V \cdot {\rm Rel}_V \Longrightarrow {\rm Id}$.
So this adjunction is a reflection with
${\rm Term}_V \morph {\bf Rel}_V \longrightarrow {\bf Term}_V$
embedding ${\bf Rel}_V$ as a subcategory of ${\bf Term}_V$, and
${\rm Rel}_V \morph {\bf Term}_V \longrightarrow {\bf Rel}_V$
reflecting ${\bf Term}_V$ into its ``subcategory'' ${\bf Rel}_V$.

Let ${\bf E} \morph {\bf \Omega} \longrightarrow {\bf adj}$,
or equivalently,
$P_{\scriptbf{E}} \morph {\bf E} \longrightarrow {\bf \Omega}$ be any dialectical base.
A {\em distributed} {\bf E}-{\em type} ${\cal C}$ is a pair ${\cal C} = \pair{{\bf C}}{|\:|_{\cal C}}$
where ${\bf C}$ is a category and $|\:|_{\cal C} \morph {\bf C} \longrightarrow {\bf \Omega}$ is a functor called an {\bf E}-{\em typing}.
Given a distributed {\bf E}-type ${\cal C} = \pair{{\bf C}}{|\:|_{\cal C}}$,
a {\em distributed} {\bf E}-{\em entity} $\Phi$ of type ${\cal C}$
is a functor $\Phi \morph {\bf C} \longrightarrow {\bf E}$ satisfying $\Phi \cdot P_{\scriptbf{E}} = |\:|_{\cal C}$.
Let ${\bf E}^{\cal C}$ denote the collection (bicomplete preorder) of all distributed {\bf E}-entities of type ${\cal C}$.
A {\em distributed} {\bf E}-{\em term} (or an {\bf E}-{\em termed distributor}) ${\cal R}$ is a pair ${\cal R} = \pair{{\bf R}}{|\:|_{\cal R}}$
where ${\bf B} \stackrel{\scriptbf{R}}{\rightharpoondown} {\bf A}$ is a distributor
and ${\bf R} \stackrel{|\:|_{\cal R}}{\Longrightarrow} {\bf \Omega}$ is a morphism of distributors called an {\bf E}-{\em terming}.
Let ${\bf Term}_E$ denote the category of distributed {\bf E}-terms and their morphisms.
Any distributed {\bf E}-term
${\cal R} = \pair{{\bf R}}{|\:|_{\cal R}}$
with distributor ${\bf B} \stackrel{\scriptbf{R}}{\rightharpoondown} {\bf A}$
and {\bf E}-{\em terming} $|\:|_{\cal R} \morph {\bf R} \longrightarrow {\bf \Omega}$
defines by composition the {\bf E}-morphism
${\bf E}^{{\cal B}} \! \stackrel{\directflow_{\cal R}}{\longrightarrow} {\bf E}^{{\cal A}}$
called {\em direct flow} (or {\em yang}),
where
$\directflow_{\cal R}$ is defined by
\[ \directflow_{\cal R}\!(\Psi)(A)
= \bigvee_{B \in |\scriptbf{B}|,e \in \scriptbf{R}[B,A]} {\bf E}^{|e|_{\cal R}}(\Psi(B)) \]
for any distributed {\bf E}-entity $\Psi \memberof {\bf E}^{{\cal B}}$ of distributed {\bf E}-type ${\cal B} = \pair{{\bf B}}{|\:|_{\cal B}}$ and any object $A \memberof {\rm Obj}({\bf A})$,
where $\bigvee$ denotes supremum,
or colimit,
in ${\bf E}(|A|_{\cal A})$.
The {\bf E}-morphism $\directflow_{\cal R}$ has as a right adjoint the {\bf E}-morphism
${\bf E}^{{\cal B}} \! \stackrel{\inverseflow_{\cal R}}{\longleftarrow} {\bf E}^{{\cal A}}$
called {\em inverse flow} (or {\em yin}),
and defined by
\[ \inverseflow_{\cal R}\!(\Phi)(B)
= \bigwedge_{A \in |\scriptbf{A}|,e \in \scriptbf{R}[B,A]} {\bf E}_{|e|_{\cal R}}(\Phi(A)) \]
for any distributed {\bf E}-entity $\Phi \memberof {\bf E}^{{\cal A}}$ of distributed {\bf E}-type ${\cal A} = \pair{{\bf A}}{|\:|_{\cal A}}$ and any object $B \memberof {\rm Obj}({\bf B})$,
where $\bigwedge$ denotes infimum,
which is limit,
in ${\bf E}(|B|_{\cal B})$.
Parallel pairs of distributed {\bf E}-terms specify
``the dialectical motion (flow, development) of entities''.
\begin{Fact}
Direct flow is left adjoint to inverse flow
$(\directflow_{\cal R} \dashv \inverseflow_{\cal R})$
for any distributed {\bf E}-term ${\cal R}$.
\end{Fact}
A parallel pair
$\tau = \parpair{{\cal B}}{\footnotecal{O}}{\footnotecal{I}}{{\cal A}}$
of distributed {\bf E}-terms is called a {\em dialectical flow specifier}.
Given a flow specifier $\tau$,
the {\em dialectical flow} (or {\em yinyang}) specified by $\tau$ is the composition
\[ \yinyang_\tau = {\bf E}^{\cal A} \stackrel{\inverseflow_{\cal I}}{\longrightarrow} {\bf E}^{\cal B} \stackrel{\directflow_{\cal O}}{\longrightarrow} {\bf E}^{\cal A} .\]
We say that $\tau$ {\em reproduces} the distributed {\bf E}-entity $\Phi \memberof {\bf E}^{\cal A}$ when $\Phi$ is a fixpoint solution of the recursive equation
$\chi \equiv \yinyang_\tau(\chi)$.
An {\em elementary dialectical {\bf E}-system}
{\sf S} is a quadruple ${\sf S} = \quadruple{T}{{\cal S}}{{\cal I}}{{\cal O}}$ consisting of:
a set (of transition symbols) $T$,
two distributed {\bf E}-types (of sites) ${\cal S} \!=\! \pair{{\cal S}_0}{{\cal S}_1}$,
an inverse flow assignment ${\cal I} \morph T \longrightarrow {\bf Term}_E[{\cal S}_0,{\cal S}_1]$, and
a direct flow assignment $O \morph T \longrightarrow {\bf Term}_E[{\cal S}_0,{\cal S}_1]$,
where ${\bf Term}_E[{\cal S}_0,{\cal S}_1]$ is the collection
of all distributed {\bf E}-terms between the distributed {\bf E}-types of sites ${\cal S}_0$ and ${\cal S}_1$.
For fixed category of transition symbols {\bf T},
we can define morphisms of dialectical {\bf E}-systems,
analogous to those for nets.
Then, for fixed ${\bf T}$,
dialectical {\bf E}-systems and their morphisms form a category
${\bf kosmos}_{\scriptbf E}^{\scriptbf T}$,
the ${\bf T}$-th fiber of
the {\em category of dialectical {\bf E}-systems} 
${\bf kosmos}_{\scriptbf E}$
(the {\bf E}-th kosmos).

If {\bf V} is any closed preorder with associated dialectical base
${\bf V} \morph \bullet_V \longrightarrow {\bf adj}$,
or equivalently $P_{\scriptbf{V}} \morph {\bf V} \longrightarrow \bullet_V$,
then dialectical {\bf V}-systems (as shown above) are (or more precisely, can be reflected into) dialectical {\bf V}-nets.
If
$D \cdot {\bf P} \morph {\bf T}_\Sigma^{\rm op} \rightarrow {\bf Set} \rightarrow {\bf adj}$
is the dialectical base,
where $\Sigma$ is an algebraic signature with term category ${\bf T}_\Sigma$,
$D$ is a $\Sigma$-algebra in functorial form,
and ${\bf P} \morph {\bf Set} \longrightarrow {\bf adj}$ is the direct-image/inverse-image dialectical base,
then dialectical $D \cdot {\bf P}$-systems are Horn clause logic programs.
Horn clause logic programs can be enriched by replacing the pseudofunctor
${\bf P} \morph {\bf Set} \longrightarrow {\bf adj}$
with the existential-Kan-quantification/substitution pseudofunctor
${\bf P}_{\scriptbf{V}} \morph {\bf Space}_V \longrightarrow {\bf adj}$,
where ${\bf P}_{\scriptbf{V}}({\cal X} \stackrel{f}{\rightarrow} {\cal Y})
= \parpair{{\bf V}^{{\cal X}}}{\subsupsize{\exists_f}}{\subsupsize{{\bf V}^f}}{{\bf V}^{{\cal Y}}}$,
and defining the dialectical base to be
$D \cdot {\bf P}_{\scriptbf{V}} \morph {\bf T}_\Sigma^{\rm op} \rightarrow {\bf Space}_V \rightarrow {\bf adj}$,
where $D$ is any {\bf V}-enriched $\Sigma$-algebra
(see \cite{Kent87b} for further development of this case).
In particular,
the special case of natural numbers ${\bf V} = {\bf N}$ enriches Horn clause logic programs with multiplicities,
and gives a proper formulation for ``predicate/transition nets'',
which are not just nets,
but full-fledged dialectical $D \cdot {\bf P}_{\scriptbf{N}}$-systems. \doublespace




\end{document}